\documentclass[12pt]{iopart}

\usepackage{iopams}  
\begin{document}

\title[Laplace-Beltrami Formalism for GR]{The Heuristic Approach to General Relativity in the Laplace-Beltrami Formalism}

\author{Noah M. MacKay
\\ ORCID ID: 0000-0001-6625-2321}

\address{Institut für Physik und Astronomie, Universit\"at Potsdam,\\ Karl-Liebknecht-Str. 24/25, 14476 Potsdam, Germany}
\ead{noah.mackay@uni-potsdam.de}
\vspace{10pt}
\begin{indented}
\item[]\today
\end{indented}

\begin{abstract}
The Laplace-Beltrami formalism, in which the Ricci tensor in the Einstein field equations (EFEs) is formulated at leading-order in terms of the partial-differential Laplace-Beltrami operator, was previously applied to coalescing compact binaries (CCBs) generating gravitational waves (GWs). Supposing that the CCB is an effective singular body -- a hollow mass-shell -- that follows a Kerr metric Ansatz, the EFEs were approached variationally such that the Ansatz geometric signature dictates the energetic output via $G_{\mu\nu}=8\pi GT_{\mu\nu}$. For the CCB mass-shell representation, the generated GW energy is treated as radiated surface energy via $E:=T_{00}V$. This surface energy yielded a close approximation to the cataloged GW coalescence energy, as previously shown in past comparisons. Given this success, it is logical to ask whether the Laplace-Beltrami formalism can be applied to other general relativistic systems, whether ``simple" or ``perturbative", beyond CCBs. 

This heuristic work focuses broadly on the EFEs themselves under the Laplace-Beltrami formalism, considering all differential orders up to second-order. This namely includes a deeper analysis on the variational methodology employed on the EFEs in the second-order sector, utilized in previous works, and the benchmark analysis of the lower first- and zeroth-order terms. This all-order report utilizes representative examples and select metric Ans\"atze to explore the formalism's practicality and its limitations; this is shown that the first-order decomposition showcases heuristically the mechanics of vector and scalar fields upon a curved spacetime.
\end{abstract}

%
\noindent{\it Keywords}: General relativity, Laplace-Beltrami formalism\\
%
\submitto{\CQG}
%
\maketitle
%
%


\section{Introduction} \label{sec1}

In November 1915, Albert Einstein derived his field equations to general relativity, which double as the field equations to gravitation \cite{Einstein:1915ca}. While the motivation was to, namely, generalize his 1905 special relativity theory to account for forms of acceleration, general relativity was able to explain much of the discrepancies in astronomy at the time -- those which Newtonian gravity had either failed to reconcile or considered impossible, perhaps nonsensical. These include the precession in Mercury's perihelion \cite{Einstein:1915bz}, and the bending of starlight by intermediating gravitational signatures (e.g. the Sun's) between the source and the observer. The latter notion was strengthened by Arthur Eddington's 1919 solar eclipse images \cite{Dyson:1920cwa}. These solutions required the novel-at-the-time picture of gravitation as the inherent curvature in spacetime sourced by resting astronomical bodies. The instantaneous strength of gravity of a local body, as measured by outside observers, is therefore gauged by the instantaneous deepness of the spacetime potential well that body generates.

The Einstein field equations (EFEs) describe the gravitational interaction as the spacetime curvature caused by energy-momentum sources, and they read as follows (using $c=1$):
\begin{equation}\label{efes}
R_{\mu\nu}-\frac{1}{2}Rg_{\mu\nu}+\Lambda g_{\mu\nu}=:G_{\mu\nu}+\Lambda g_{\mu\nu}=8\pi GT_{\mu\nu}.
\end{equation}
The left-hand side encodes the warped geometric response sourced by the energy-momentum source $T_{\mu\nu}$, in the shorthand known as the Einstein tensor $G_{\mu\nu}$. The Einstein tensor is linearly constructed from the metric $g_{\mu\nu}$ and Ricci curvature $R_{\mu\nu}$ tensors, and the Ricci trace scalar $R=g^{\mu\nu}R_{\mu\nu}$, where $g^{\mu\nu}$ is the inverse metric tensor. For completeness, the Einstein tensor is accompanied by the additive cosmological term $\Lambda g_{\mu\nu}$, which was originally introduced by Einstein to permit a static universe. While the static universe notion was disproven by Edwin Hubble's discovery of an expanding universe \cite{Hubble:1926yw, Hubble:1929ig}, $\Lambda$ remains a central component of modern cosmology, in particular within the standard $\Lambda\mathrm{CDM}$ model despite discrepancies such as the Hubble Tension. However, more local systems often neglect this contribution; in this work we neglect $\Lambda g_{\mu\nu}$. 

Systematically, the EFEs are a collection of coupled, nonlinear equations. The geometry embedded in $G_{\mu\nu}$ depends on (besides the metric and Ricci tensors) relevant Christoffel symbols $\Gamma^\alpha_{\mu\nu}$, and one computes $G_{\mu\nu}$ generally by directly following the standard hierarchy in the EFEs (i.e. meticulously following the chain $g_{\mu\nu}\rightarrow g^{\mu\nu}\rightarrow\Gamma^\alpha_{\mu\nu}\rightarrow R_{\mu\nu}\rightarrow R\rightarrow G_{\mu\nu}$). Known solutions to the EFEs, e.g. Schwarzschild \cite{Schwarzschild:1916uq, Schwarzschild:1916ae}, Tolman-Oppenheimer-Volkoff (TOV) \cite{Tolman:1939jz, Oppenheimer:1939ne}, and Kerr \cite{Kerr:1963ud, Boyer:1967, Chandrasekhar:1985kt}, were obtained via this meticulous chain by first assigning a well-defined energy-momentum source for $T_{\mu\nu}$: either vacuum $T_{\mu\nu}=0$ \cite{Schwarzschild:1916uq, Schwarzschild:1916ae, Kerr:1963ud, Boyer:1967} or perfect fluid $T_{\mu\nu}=(\epsilon+p)u_\mu u_\nu+pg_{\mu\nu}$ \cite{Tolman:1939jz, Oppenheimer:1939ne}. Other solutions to the EFEs include Kerr-Newman for a rotating, charged mass \cite{Newman:1965tw, Newman:1965my}, Vaidya for a stationary mass either emitting or absorbing dusts \cite{Poisson:2004, Griffiths:2009}, and Friedmann-Lemaitre-Robertson-Walker (FLRW) for a homonegeous and isotropic universe described as a perfect fluid \cite{Friedmann:1924bb, Lemaitre:1927zz, Robertson:1933zz, Walker:1937qxv}.

The exact treatment of general relativity works best for ``simple" astrophysical systems, mainly pertaining isolated, static / rotating bodies. These include e.g. black holes (BHs) \cite{Schwarzschild:1916uq, Schwarzschild:1916ae, Kerr:1963ud, Boyer:1967}, with more detailed analyses since 1916 having involved precise techniques such as differential topology \cite{Penrose:1964wq} and modern contemporary approaches (e.g. black branes and fuzzballs in string theory), and neutron stars (NSs) \cite{Tolman:1939jz, Oppenheimer:1939ne}, with added rotations described by the Hartle-Thorne metric \cite{Hartle:1967he, Hartle:1968si} and tidal distortions caused by Regge-Wheeler gauged metric perturbations \cite{Regge:1957td, Thorne:1997kt}. However, more generally, an exact analyical treatment of the EFEs (i.e. following the streamlined hierarchy) becomes intractable for more complicated, ``perturbative" astrophysical systems. This has motivated several complementary approaches, such as the 3+1 decomposition in numerical relativity via the Arnowitt-Deser-Misner (ADM) formalism \cite{Arnowitt:1962hi} and heuristic analysis of spacelike singularities through the Belinksi-Khalatnikov-Lifshitz (BKL) conjecture \cite{Belinski:1971, Belinskii:1972sg, Belinskii:1973sud, BK, Belinsky:1982pk, Belinski:2017fas}.

One example of such a perturbative system described by the EFEs is (rather famously) two-body, quadrupolar gravitational wave (GW) sources, which in nature are well understood to be coalescing compact binaries (CCBs). This coalescence complicates the otherwise simple, circular GW-forms to incorporate dynamic post-Newtonian (PN) \cite{Blanchet:2013haa} and perturbative post-Minkowskian (PM) \cite{Damour:2016gwp} corrections. Past the PN regime, CCBs undergo the inspiral-merger-ringdown (IMR) process, where the waveform intensifies with a dynamic frequency and amplitude enhancement, until reaching the maximum peak at the coalescence time $t_C$. Past coalescence, ignoring tidal deformations, the waveform dampens and stiffens into a zero flat-line. 

A very successful, state-of-the-art procedure to describe CCBs (semi-)analytically is the effective one-body (EOB) model \cite{Buonanno:1998gg}, namely mapping CCBs onto an effective one-body problem and essentially depicting inspiral with a ``marble in a funnel" picture. The associating energetics is described by the EOB Hamiltonian \cite{Buonanno:1998gg, Damour:2009zoi, Damour:2012mv}, which is extractible conventionally from the EFEs. However, the EOB picture particularily relies on calibration with numerical relativity and input source parameters deduced by Bayesian analysis of detected signals. Despite the accuracy generated by the EOB Hamiltonian and the framework's integration into GW analysis, computing the Hamiltonian's energy eigenvalue and the respective waveform families is computationally (and temporally) expensive due to the inherent complexities of CCBs. 

Returning to the subject of geometric analyses, the independent analysis by Chow and Knopf \cite{Chow:2004} (i.e., their lemma 3.32) suggests that, in appropriate coordinate settings, the Ricci tensor may be expressed schematically as the partial-differential Laplace-Beltrami operator acting on the metric, supplemented by lower-order terms:
\begin{equation}
R_{\mu\nu}\simeq-\frac{1}{2}\Delta^\mathrm{LB} g_{\mu\nu}+\left(\mathrm{lower~order~terms}\right).
\end{equation} 
The operator $\Delta^\mathrm{LB}$ is the Laplace-Beltrami operator, and when the metric is well-defined and has a non-zero determinant, it can be written in a coordinate expression (i.e. in the Christoffel symbol-free form) as:
\begin{equation}\label{lbo}
\Delta^\mathrm{LB}=\frac{1}{\sqrt{-g}}\,\partial_\alpha\left(\sqrt{-g}g^{\alpha\beta}\partial_\beta \right),
\end{equation}
where $\sqrt{-g}=\sqrt{-\mathrm{det}(g_{\mu\nu})}$. This observation has motivated a Laplace-Beltrami-based reformulation of the EFEs, where the Ricci tensor is expressed at face-value by this operator. This notion was explored in the context of CCBs in Ref. \cite{MacKay:2024qxj} and expanded upon in Ref. \cite{MacKay:2025uyg}.  In this context, the geometric content of the theory is recast in terms of differential operators acting directly on the metric components. Thus, one adopts a methodology analogous to the quantum-mechanical variational method to approach the EFEs: 
\begin{enumerate}
\item A physically sensible metric Ansatz $g_{\mu\nu}$ (a known solution to the EFEs) under a relevant choice of coordinates are inserted into the Laplace-Beltrami operator, uniquely defining it. 
\item Each component of the same metric Ansatz is fed into the uniquely-defined operator, extracting leading-order geometric contributions that define the corresponding (effective) Ricci tensor component. In this manner, the number of non-zero components in the effective $R_{\mu\nu}$ is at most the number of components in the metric Ansatz tensor. 
\item These can then be interpreted as the corresponding (effective) energy-momentum content, constructing $T_{\mu\nu}$ component-wise. One must be aware that such identifications are effective descriptions and approximations, rather than as exact tensorial relations. 
\end{enumerate}
However, the new hierarchal chain of the EFEs in this formalism is slightly more streamlined, $g_{\mu\nu}\rightarrow g^{\mu\nu}\rightarrow R_{\mu\nu}\rightarrow R\rightarrow G_{\mu\nu}\rightarrow T_{\mu\nu}$. This completely bypasses the Christofel symbols, such that the inherent first-differential order behavior of $\Gamma^\alpha_{\mu\nu}$ is readily nestled inside the Laplace-Beltrami operator.

For e.g. CCBs approached in \cite{MacKay:2024qxj, MacKay:2025uyg}, the effective one body picture was reinterpreted from the ``marble in the funnel" perspective to a  compact-body mass-shell: a Kerr-like body with a constant reduced mass measure, a shrinking diameter and increasing rotational velocity across coalescence. Coalescence ends at the ``innermost" shell radius of $\rho=2GM$, and the shell's surface energy eigenvalue encoded as $T_{00}$ can approximate the energy released as GWs. Given the mass-shell model qualitatively mimics a Kerr-like object, it follows the Kerr metric Ansatz. Following the new hierarchal chain through the Laplace-Beltrami formulation of the Ricci tensor, one yields c.f. \cite{MacKay:2024qxj, MacKay:2025uyg} the coalescence energy eigenvalue of
\begin{equation}\label{gwen}
E(t_C)\simeq 0.826\,\nu \mu \left(1-5.276 \beta_C^2 \right),
\end{equation}
where $\nu:=\mu/M$ is the reduced mass ratio and $\beta_C$ is the speed ratio depending on the normalized rotation speed of the CCB at coalescence (i.e., at the moment of merger). This energy eigenvalue shows that $E\propto\nu\mu$, and the above formula for coalescence energy has so far agreed well with the energy radiated from noteworthy GW events cataloged in Refs. \cite{GWOSC, LIGOScientific:2018mvr, LIGOScientific:2021usb, KAGRA:2021vkt, LIGOScientific:2025slb}.  It is the success of this formalism for CCBs and the alignment between model predictions and observation that has motivated this current study.

Provided the moderate success of the Laplace-Beltrami formalism when applied to this CCB mass-shell model, a natural question is whether this formalism can be applied to other general relativistic systems. After the introduction of the formalism in Section \ref{sect:lb}, we look broadly at the EFEs themselves, while applying Ansatz metrics with hopes to recover conventional insight as well as offer new information. After the analysis, we make concluding statements in Section \ref{sect:concl} with a scope of the possible directions to take this formalism.

\section{The Laplace-Beltrami Formalism}\label{sect:lb}

In perturbative quantum mechanics, the variational method estimates the energy eigenvalue of an impossible, perturbative Hamiltonian by assuming an Ansatz wavefunction, computing the expectation value of the Hamiltonian, and minimizing it to approximate the energy at the ground state. For complicated general relativistic systems whereby their energetics are inherently perturbative, we adopt a similar philosophy. We begin with a well-defined metric, under a choice of coordinates, with a non-zero determinant as our Ansatz, and apply differential operations that are inherent in the Christoffel symbols and the Ricci tensor. Using the EFEs, we construct the associated energy-momentum tensor $T_{\mu\nu}$ component-wise, such as the energy density $T_{00}$ that embeds the energy eigenvalue $E=T_{00}V$.

Schematically, the Ricci tensor obeys $R\sim\partial\Gamma+\Gamma\Gamma$ and the Christoffel symbols satisfy $\Gamma\sim g^{-1}\partial g$. Thus, we claim that the Ricci tensor behaves qualitatively like a Laplacian operator acting on the metric tensor: $R\sim \nabla(g^{-1}\partial g)\sim\nabla^2 g$, where $\nabla\sim\partial+\Gamma$ is the covariant derivative. This interpretation is consistent with lemma 3.32 from Chow and Knopf \cite{Chow:2004}, which expresses the Ricci tensor as a covariant Laplacian, followed by lower-order terms defined here:
\begin{equation}\label{ricci}
R_{\mu\nu}\simeq-\frac{1}{2}\Delta^\mathrm{LB} g_{\mu\nu}+\frac{1}{2}\nabla_{(\mu}V_{\nu)}+A_{\mu\nu},
\end{equation} 
The operator $\Delta^\mathrm{LB}$ is defined as Eq. (\ref{lbo}), whose form is distinctly coordinate-compatible. Specific expressions for the operator are therefore dependent on the choice of coordinates in the metric Ansatz.

The lower-order terms include respectively a first-order covariant derivative of a rank-1 vector $V_\nu$ (where $\nabla_{(\mu}V_{\nu)}=\nabla_\mu V_\nu+\nabla_\nu V_\mu$ is a symmetric combination) and a zeroth-order auxiliary rank-2 tensor $A_{\mu\nu}$. It should be noted that Appendix A in Ref. \cite{MacKay:2024qxj} introduces the first-order term without the $1/2$ scaling. In this work, this scaling is introduced, as it would prove essential in the respective first-order analysis and to correct any unnecessary overcounting. In the Christoffel symbol-free form motivated by $\Delta^\mathrm{LB}$, the covariant gradient of a rank-$n$ trial tensor $f^{a_1\dots a_n}_{b_1\dots b_n}$ can be defined as
\begin{equation}
\nabla_\mu f^{a_1\dots a_n}_{b_1\dots b_n}=\frac{1}{\sqrt{-g}}\partial^\nu \left(\sqrt{-g}g_{\mu\nu}f_{a_1\dots a_n}^{b_1\dots b_n}\right).
\end{equation}
Note that the metric tensor is introduced in the parentheses above, which comes at the cost of swapped indices for the trial tensor $f^{a_1\dots a_n}_{b_1\dots b_n}$. This revision enforces the convention of metric compatibility, should $f_{a_1\dots a_n}^{b_1\dots b_n}=g^{\mu\nu}$ (i.e., $f^{a_1\dots a_n}_{b_1\dots b_n}=g_{\mu\nu}$), whereby $\nabla_\mu g^{\mu\nu}=\nabla_\mu g_{\mu\nu}=0$. This is given by $g_{\mu\nu}g^{\mu\nu}\in\mathbb{N}$ (i.e., 4 for a $3+1$ spacetime) and $g_{\mu\alpha}g^{\mu\beta}=\delta_\alpha^{~\beta}$, whereby the coordinate-based derivative of an integer is zero. This is independent on the choice of coordinates. 

One should note that, while $\nabla_\mu g^{\mu\nu}=0$ is enforced for first-order covariant gradients, the second-order Laplacian is purely coordinate-dependent and nontrivial for individual metric tensor components. Even for metric Ans\"atze where $R_{\mu\nu}=0$ conventionally, the second-order Laplace-Beltrami formalism allows for non-zero, effective values in $R_{\mu\nu}$, provided the components of $g_{\mu\nu}$ are distinctly coordinate dependent. This is particularly useful for approximating our desired quantites for a complicated system, constructing its effective energy-momentum tensor component-wise. 

The corresponding Ricci scalar in the Laplace-Beltrami formalism is
\begin{equation}\label{rscale}
R=-\frac{1}{2}g^{\mu\nu}\Delta^\mathrm{LB} g_{\mu\nu}+\nabla^\mu V_\mu +A,
\end{equation}
implying $g^{\mu\nu}\nabla_\mu=\nabla^\nu$ and $g^{\mu\nu}A_{\mu\nu}=A$. This form, being a scalar, is manifestly coordinate-independent, which enables us the freedom to choose a particular metric Ansatz and choice of coordinates that best fits our particular scenario. By implementing Eqs. (\ref{ricci}) and (\ref{rscale}) into the Einstein tensor and neglecting $\Lambda g_{\mu\nu}$, we arrive at a natural decomposition of $G_{\mu\nu}$ in the Laplace-Beltrami formalism:
\numparts
\begin{eqnarray}\label{2ndord}
&\mathrm{2nd-Order:}\quad G_{\mu\nu}^{(2)}=-\frac{1}{2}\Delta^\mathrm{LB} g_{\mu\nu}+\frac{1}{4}\left(g^{\mu\nu}\Delta^\mathrm{LB} g_{\mu\nu}\right)g_{\mu\nu},\\\label{1stord}
&\mathrm{1st-Order:}\quad G_{\mu\nu}^{(1)}=\frac{1}{2}\nabla_{(\mu}V_{\nu)}-\frac{1}{2}\left(\nabla^\mu V_\mu\right)g_{\mu\nu},\\\label{0thord}
&\mathrm{0th-Order:}\quad G_{\mu\nu}^{(0)}=A_{\mu\nu}-\frac{1}{2}Ag_{\mu\nu},\\\label{efes2}
&G_{\mu\nu}=8\pi GT_{\mu\nu}\quad\Rightarrow \quad\sum_{i=0}^2G^{(i)}_{\mu\nu}=8\pi G\sum_{i=0}^2T^{(i)}_{\mu\nu}.
\end{eqnarray}
\endnumparts

\subsection{Bianchi Identity Sanity Check}

A natural verification of the formalism is upholding the Bianchi identities and the concurrent energy-momentum conservation. Given $G_{\mu\nu}\propto T_{\mu\nu}$ via Eq. (\ref{efes}) as well as Eq. (\ref{efes2}), standard general relativity enforces the vanishing covariant derivative on both sides of the EFEs. Through $\nabla^\mu G_{\mu\nu}=0$, one has the reduced Bianchi identity of the form
\begin{equation}
\nabla^\mu R_{\mu\nu}=\frac{1}{2}\nabla_\nu R,
\end{equation} 
with $\nabla^\mu g_{\mu\nu}=0$ readily implied. To verify that the reduced Bianchi identity remains satisfied within this Laplace-Beltrami formalism, we must briefly address the leading second-order part of the decomposition. It should be noted that the Laplace-Beltrami operator acts explicitly on our choice of Ansatz metric component-wise in a coordinate-dependent form, provided as Eq. (\ref{lbo}), which can therefore provide nontrivial results. In this sanity check, the Laplacian operator is treated in its coordinate-independent form $\Delta^\mathrm{LB}:=\nabla_\alpha\nabla^\alpha$ to verify general covariance and metric compatibility. By this caveat, the leading second-order term naturally vanishes. Therefore, only the lower-order terms remain; equating both sides of the Bianchi identity therefore gives the following:
\begin{eqnarray}
&\frac{1}{2}\nabla^\mu\nabla_{(\mu}V_{\nu)}+\nabla^\mu A_{\mu\nu}=\frac{1}{2}\left(\nabla_\nu\nabla^\mu V_\mu +\nabla_\nu A \right)\\\nonumber
&\Rightarrow\quad \nabla^\mu\nabla_{(\mu}V_{\nu)}=\nabla_\nu\nabla^\mu V_\mu\quad\mathrm{and}\quad\nabla^\mu A_{\mu\nu}=\frac{1}{2}\nabla_\nu A.
\end{eqnarray}

One can see readily that the Bianchi identity is satisfied for the zeroth-order auxiliary tensor $A_{\mu\nu}$. For the first-order term, one can utilize the symmetric combination to yield an emergent Laplace-Beltrami operator acting on the 4-vector:
\begin{equation}
\nabla^\mu \nabla_\mu V_\nu:=\Delta^\mathrm{LB}V_\nu=\nabla_\nu\nabla^\mu V_\mu-\nabla^\mu\nabla_\nu V_\mu.
\end{equation} 
The right-hand side can be expressed as a commutator of covariant derivatives acting on $V_\mu$:
\begin{equation}
\nabla_\nu\nabla^\mu V_\mu-\nabla^\mu\nabla_\nu V_\mu=g^{\mu\rho}[\nabla_\nu,\nabla_\rho]V_\mu=g^{\mu\rho}R_{\sigma\mu\nu\rho}V^\sigma,
\end{equation}
where the rightmost expression follows from the definition of the Riemann tensor. As $g^{\mu\rho}R_{\sigma\mu\nu\rho}$ defines the Ricci tensor $R_{\sigma\nu}$, the first-order Bianchi identity in the Laplace-Beltrami formalism gives a curvature-coupled, inhomogeneous wave-like equation for the 4-vector:
\begin{equation}\label{veceq}
\Delta^\mathrm{LB}V_\nu=R_{\sigma\nu}V^\sigma.
\end{equation}

The derivation of Eq. (\ref{veceq}) is truly a unique recovery, as it shows a background-dependent wave equation of a rank-1 vector being the product of the curvature coupling to that same vector -- equivalent to an eigenvalue expression. If $V_\nu$ were regarded as an independent dynamical quantity, this relation would describe the propagation of a spin-1 field on a curved manifold, or alternatively a natural geometric origin for an emergent rank-1 field associated with curvature. In any case, Eq. (\ref{veceq}) is an equation that holds for anyone's choice of metric Ansatz and coordinate system. As a trade-off, one's choice of metric Ansatz and coordinates ultimately influences the structure of the equation. 

E.g., for a vector field in flat vacuum, whereby $g_{\sigma\nu}=\eta_{\sigma\nu}$ and $R_{\sigma\nu}=0$ (even in the Laplace-Beltrami formalism), we recover the standard d'Alembertian wave equation with no source:
\begin{equation}\label{s1f}
\Box V_\nu=0.
\end{equation} 
This vector wave equation is the conventional equation for massless, spin-1 fields, e.g. the photon field in quantum electrodynamics and the gluon field in quantum chromodynamics. Therefore, this vector in the flat vacuum must obey the Lorentz gauge $\partial_\mu V^\mu=\partial^\mu V_\mu=0$, imposing two degrees of freedom that are transverse to the axis of propagation.

Given this massless form, mass emerges otherwise on the right-hand side of Eq. (\ref{s1f}). As it is seen in Eq. (\ref{veceq}), the respective right-hand side is the same vector field coupled to the Ricci curvature tensor. Therefore, in this first-order Laplace-Beltrami formalism, the emergence of mass is associated with the presence of a non-zero Ricci contribution. This upholds the conventional understanding of Ricci curvature signatures being sourced by localized mass. This Ricci contribution is either conventionally obtained via the EFEs or effectively computed in the second-order Laplace-Beltrami formalism. One must note that, if opting for the effective Laplace-Beltrami computation of $R_{\mu\nu}\propto\Delta^\mathrm{LB}g_{\mu\nu}$, the associating Laplace-Beltrami operator must be consistent with the one applied on the vector field. 

On the other hand, to address Eq. (\ref{veceq}) upon a curved vacuum, where the Laplace-Beltrami operator depends on our choice of metric Ansatz, our approach to treat $R_{\sigma\nu}$ on the right-hand side rests in either: (i) $R_{\sigma\nu}=0$ in following convention of (curved) vacuum spacetimes, (ii) specifically introduce components of $R_{\sigma\nu}\propto\Delta^\mathrm{LB}g_{\sigma\nu}$ via the 2nd-order equation to solve for $V_\nu$ component-wise, or (iii) effectively describing $R_{\sigma\nu}$ in terms of the Weyl curvature $C_{\sigma\mu\nu\rho}$ that energy directly warps. In any case, the curved-background wave equation reads as an eigenvalue equation, depending on the avenue we choose among (i)--(iii). 

\section{Second-Order Laplace-Beltrami}

One observes from Eq. (\ref{2ndord}) that $G_{\mu\nu}\propto g_{\mu\nu}$ (dropping the Latin index for the order), with the Laplace-Beltrami operator and Ricci scalar specifically defined given our choice of Ansatz metric. E.g., when finding the energy eigenvalue of our given system, $G_{00}\propto g_{00}$ for any choice of metric Ansatz, whereby we determine $T_{00}=G_{00}/(8\pi G)$ as well as $E=T_{00}V$. For instance, given a CCB viewed from a luminosity distance $D\gg L$, if we assume it to be a spinning, compact object, our choice of Ansatz metric is the Kerr metric \cite{Kerr:1963ud, Boyer:1967, Chandrasekhar:1985kt} in Boyer-Lindquist coordinates. This was done in Refs. \cite{MacKay:2024qxj, MacKay:2025uyg} to derive the surface energy of this Kerr-like CCB as the approximated emitted GW energy, in Ref. \cite{MacKay:2025uyg} to specifically derive Eq. (\ref{gwen}). In this work, we repeat the methods previously used, but we impose various scenarios and new metric Ans\"atze as a proof of concept. This is to test both the useful applications of the formalism and the limitations where the formalism may not be sensible to use.

\subsection{On One's Choice of Coordinates}

Given the coordinate-dependence in the Laplace-Beltrami operator via Eq. (\ref{lbo}), the definitions of both the Ricci tensor in this formalism $R_{\mu\nu}\propto\Delta^\mathrm{LB}g_{\mu\nu}$ and the Ricci scalar heuristically constructed as $R\propto g^{\mu\nu}(\Delta^\mathrm{LB}g_{\mu\nu})$ are ultimately dependent on the choice of coordinates. The choice of coordinates, for many cases, comes hand-in-hand with the underlying choice of metric Ansatz. E.g., the Schwarzschild and TOV metrics are written under spherical coordinates, describing a spherical body, and e.g. the Minkowski and flat FLRW metrics are written under Cartesian coordinates. It is important to note the impact of coordinate systems in this Lapalce-Beltrami formalism, as certain choice of coordinates would lead to unphysical divergences, such as those previously encountered in Refs. \cite{MacKay:2024qxj, MacKay:2025uyg} under Boyer-Lindquist coordinates (based on spherical coordinates).

Provided e.g. a general yet diagonal Cartesian- and spherical-coordinated metric:
\numparts
\begin{eqnarray}
&\mathrm{Cartesian:}\quad ds_\mathrm{Cart.}^2=g_{00}dt^2+g_{11}dx^2+g_{22}dy^2+g_{33}dz^2,\\
&\mathrm{Spherical:}\quad ds_\mathrm{Sph.}^2=g_{00}dt^2+g_{11}dr^2+g_{22}d\theta^2+g_{33}d\phi^2,
\end{eqnarray}
\endnumparts
the Laplace-Beltrami operator $\Delta^\mathrm{LB}$ is expanded as follows:
\begin{eqnarray}
\Delta^\mathrm{LB}=\frac{1}{\sqrt{-g}}\Bigg[&\partial_{x_0} \Big(\sqrt{-g}g^{00}\partial_{x_0}\Big) + \partial_{x_1} \Big(\sqrt{-g}g^{11}\partial_{x_1}\Big)\\\nonumber
& + \partial_{x_2} \Big( \sqrt{-g}g^{22}\partial_{x_2}\Big) + \partial_{x_3} \Big(\sqrt{-g}g^{33}\partial_{x_3} \Big) \Bigg].
\end{eqnarray}
Here, $(x_0, x_1, x_2, x_3)$ is the coordinate system used in the metric, i.e. $(t,x,y,z)$ for Cartesian or $(t,r,\theta,\phi)$ for spherical, and $\sqrt{-g}=\sqrt{-\mathrm{det}(g_{\mu\nu})}$ is metric-specific and typically coordinate-dependent. The Ricci tensor $R_{\mu\nu}$ in the Lapalce-Beltrami formalism is constructed component-wise, by feeding the coordinate-dependent metric components $g_{ii},\,i=0,1,2,3$ into the operator and evaluating the relevant coordinate derivatives. The Ricci scalar $R=g^{\mu\nu}R_{\mu\nu}$  is simply the trace:
\begin{eqnarray}
R\propto\sum_{i=0}^3 \frac{g^{ii}}{\sqrt{-g}}\Bigg[&\partial_{x_0} \Big(\sqrt{-g}g^{00}\partial_{x_0}\Big) + \partial_{x_1} \Big(\sqrt{-g}g^{11}\partial_{x_1}\Big)\\\nonumber
& + \partial_{x_2} \Big( \sqrt{-g}g^{22}\partial_{x_2}\Big) + \partial_{x_3} \Big(\sqrt{-g}g^{33}\partial_{x_3} \Big) \Bigg]g_{ii}.
\end{eqnarray}
We see readily for e.g. the Minkowski metric: $ds^2=-dt^2+dx^2+dy^2+dz^2$, that $g_{00}=-1$ and $g_{jj}=1$ ($j=1,2,3$) obtain: (i) the d'Alembert wave operator from the Laplace-Beltrami operator, and (ii) $R_{\mu\nu}=0$ provided integer metric components. A trival Ricci tensor due to a flat metric, even in this formalism, remains true to the nature of flat spacetimes.

However, we turn our attention to spherically-coordinated metrics. As $g_{33}\propto\sin^2\theta$ typically under spherical coordinates, this leads to $g^{33}\propto\csc^2\theta$ that is present in the $g^{33}R_{33}$ contribution in the trace. In the Laplace-Beltrami formalism, i.e. as found in Ref. \cite{MacKay:2024qxj}, this contribution is divergent along the poles $\theta=0,\pi$, leading to the entire scalar value of $R$ to be divergent. One may be tempted to designate a value of $\theta$ to be e.g. $\pi/2$ to maintain convergence. However, this designation is only relevant for test-particle dynamics under the influence of the spherical source (e.g. orbital mechanics near a Schwarzschild body). Here, in this context of describing the spherical source itself, we have to consider full-angular integration, and such evaluations must be finite. One might also be tempted to omit the divergent term altogether, to ``sweep it under the rug," but this is physically nonsensical. Thus, as demonstrated in Refs.  \cite{MacKay:2024qxj, MacKay:2025uyg}, one would turn to using effective Ricci scalars as a means of analytical completeness. Options for effective Ricci scalars include e.g. the conventional Ricci scalar of the metric, and a curvature surrogate in terms of the Kretschmann scalar $K=R_{\mu\nu\alpha\beta}R^{\mu\nu\alpha\beta}$ (as pursued in \cite{MacKay:2024qxj, MacKay:2025uyg}). The need of a surrogate scalar, so far, is only presistent for spherical-coordinated metric Ans\"atze.

I remind, even for metric Ans\"atze where $R_{\mu\nu}=0$ conventionally, the second-order Laplace-Beltrami formalism allows for non-zero, effective values in $R_{\mu\nu}$, provided the components of $g_{\mu\nu}$ are distinctly coordinate dependent. This is particularly useful for approximating our desired quantites for our system of interest, constructing its effective energy-momentum tensor component-wise.

\subsection{Example I: The Schwarzschild Metric Ansatz}

Consider a resting, compact object with mass $m$ that qualitatively obeys the Schwarzschild metric \cite{Schwarzschild:1916uq, Schwarzschild:1916ae} in ingoing Eddington-Finkelstein (EF) coordinates \cite{Eddington:1924pmh, Finkelstein:1958zz} -- to avoid coordinate singularities --, written here as
\begin{equation}
ds^2=-f(r)dv^2+2dvdr+r^2d\Omega^2.
\end{equation}
Here, $d\Omega^2=d\theta^2 +\sin^2\theta d\phi$ is the solid angular element, $dv=dt+dr^*=dt+dr/f(r)$ is an ingoing horizon coordinate that removes the coordinate singularity at $r=2Gm$, and $f(r)=(1-2Gm/r)$ is a Schwarzschild potential function.

It might seem counter-intuitive to analyze BHs, or more broadly Schwarzschild bodies, in the Laplace-Beltrami formalism. However, we want to verify in this framework that, generally, the energy eigenvalue of a resting object follows the mass-energy relation $E=m$ at its surface. This is, logically, no different for a BH; provided the introductory works on Hawking radiation \cite{Hawking:1974rv, Hawking:1975vcx}, the BH's mass-energy relation at the horizon scale $r=2Gm$ is related with heat flow, enabling one to derive the formulae for BH entropy $S\propto A_\mathrm{BH}/4$ and temperature $T\propto 1/m$.

One can find the determinant of the Schwarzschild metric in EF coordinates to be $\mathrm{det}(g_{\mu\nu})=-r^4\sin^2\theta$ (exactly equal to the metric determinant under spherical coordinates), and the Ricci scalar under this metric is $R=0$. As expected, the Schwarzschild metric is a vacuum solution for static BHs, from which $R_{\mu\nu}=0$ conventionally. However, we define via Eq. (\ref{ricci}) nontrivial effective values in $R_{\mu\nu}$, particularly in the $R_{00}$ component via $g_{00}$. Conveniently, all metric elements depend only on $r$ and $\theta$, however the choice of coordinates slighly complicates the Schwarzschild-metric Laplace-Beltrami operator:
\begin{eqnarray}
\Delta^\mathrm{LB}_\mathrm{Schw.EF}=\frac{1}{\sqrt{-g}}\Big[&-\partial_v\left(\sqrt{-g}\frac{1}{f(r)}\partial_v\right)+\partial_v\left(\sqrt{-g}\partial_r\right)+\partial_r(\sqrt{-g}\partial_v)\\\nonumber
&+\frac{1}{r^2}\partial_\theta(\sqrt{-g}\partial_\theta)\Big],
\end{eqnarray}
where $\sqrt{-g}=r^2\sin\theta$. Recalling that $dv=dt+dr/f(r)$, we define $\partial_v=\partial_t+f(r)\partial_r$, and given the relevant metric entries do not depend on $t$, we write effectively $\partial_v\rightarrow f(r)\partial_r$:
\begin{equation}\label{lbef}
\Rightarrow \Delta^\mathrm{LB}_\mathrm{Schw.EF}=\frac{1}{r^2\sin\theta}\left[\sin\theta \partial_r(r^2 f(r)\partial_r)+\partial_\theta(\sin\theta\partial_\theta)\right].
\end{equation}
Using Eq. (\ref{ricci}) with Eq. (\ref{lbef}), one can straightforwardly compute $R_{00}$ to be non-zero:
\begin{equation}\label{r00}
R_{00}=\frac{2G^2m^2}{r^4}.
\end{equation}

In approaching the Ricci scalar $R$, we are presented the three choices first proposed in Ref. \cite{MacKay:2024qxj} and mentioned earlier in this work: (i) set $R=0$ in accordance to our vacuum metric Ansatz, (ii) heuristically construct $R$ using a Laplace-Beltrami trace over the metric: $R\propto g^{\mu\nu}\Delta^\mathrm{LB} g_{\mu\nu}$, or (iii) define an effective Ricci scalar by leveraging the optimal curvature content of the nonvanishing Kretschmann scalar: $K=R_{\alpha\beta\mu\nu}R^{\alpha\beta\mu\nu}$ \cite{dInverno:1992gxs}. If we adopt $R=0$, following convention of the Schwarzschild metric as our used metric Ansatz, then the energy density contribution $T_{00}\propto G_{00}$ depends entirely on the Ricci tensor component $R_{00}$. Suppose we do this, given Eq. (\ref{r00}) is a simple term, we have
\begin{equation}
T_{00}=\frac{1}{4\pi}\frac{Gm^2}{r^4}\quad\Rightarrow\quad E=T_{00}V=\frac{1}{3}\frac{Gm^2}{r},
\end{equation}
where, at the horizon radius of $r=2Gm$, we yield the surface energy of $E=m/6$. While this is a factor of $m$, it is reduced; we recover the unscaled rest mass energy $E=m$ at the radial position $r=Gm/3$, or one-sixth of the horizon radius, well inside the BH horizon. 

This demonstrates that this variational approach to the EFEs, here in the 2nd-order formalism, can only approximate well-defined systems close to our expectation. Of course, this is under the simple assumption of $R=0$, while our other options may help enhance our accuracy. This is more so the case given e.g. that Eq. (\ref{ricci}) can produce non-zero Ricci tensor entries for the Schwarzschild metric Ansatz beyond $R_{00}$ (see footnote\footnote{That is, except for $R_{11}$ in EF coordinates, as $g_{11}=0$.}). In the Kretschmann scalar case, to be specific, we propose a naive substitution of the Ricci scalar by $R\sim-\sqrt{K}$, which is motivated by dimensional consistency and by energetic considerations (see footnote\footnote{The square-root of $K$ restores dimensional consistency with $R$. The minus sign is introduced heuristically: primarily to offset the negativity of $g_{00}$ in $G_{00}$, thereby yielding a physically reasonable, positive energy density.}). 

In pursuing the second option of $R\propto g^{\mu\nu}\Delta^\mathrm{LB}g_{\mu\nu}$, there is a fundamental issue in the $g^{11}R_{11}$ contribution of the trace: $g_{11}:=g_{rr}=0$, thus imposing an explosive term via $g^{11}=1/0$. To avoid this, one might impose a constant screening parameter $g_{11}=\xi$ with the smallness condition $|\xi|\ll1$, such that we avoid explosive divergences. If this is done, provided readily $R_{11}=0$, then $g^{11}R_{11}=0$ easily, avoiding a $0/0$ paradox. Thus, the Laplace-Beltrami-formulated trace is presented as follows:
\begin{equation}
R=\frac{2G^2m^2}{r^3(2Gm-r)}-\frac{8Gm}{r^3}+\frac{5}{r^2}-\frac{2\cot^2\theta}{r^2},
\end{equation}
which, given the cotangent function via the $g^{33}R_{33}$ contribution of the trace, diverges when one performs polar-angular integration over $\theta\in[0,\pi]$. Because the Ricci scalar is, namely, a scalar, all angular terms must be integrated over the full range whenever they are present, such that the energy density $T_{00}$ is enclosed by a spherical volume. One would be tempted to omit the divergent term by ``sweeping it under the rug," but this is physically nonsensical as it is part of the trace $g^{\mu\nu}R_{\mu\nu}$.

For the Schwarzschild metric, independent on one's choice of coordinates, the Kretschmann scalar is non-zero:
\begin{equation}
K=\frac{48G^2m^2}{r^6};
\end{equation}
being that this is readily positive and the Ricci scalar contribution in the Einstein tensor is negated, proposing a heuristic modification where $R_\mathrm{eff}=-\sqrt{K}$ allows us to explore the \textit{additive} curvature-induced contributions to $G_{00}=R_{00}-R_\mathrm{eff}g_{00}/2$, and how they play a role in calculating the energy density $T_{00}$ beyond the $R_{00}$ minimum. This substitution leads to an effective expression for $G_{00}$ as an expansion in leading orders of $G$ (a PM expansion):
\begin{eqnarray}
G_{00}&=\frac{G^2m^2}{r^4}-\frac{2Gm}{r^3}f(r)\sqrt{3}\\\nonumber
&=\frac{2G^2m^2}{r^4}+\frac{4\sqrt{3}G^2m^2}{r^4}-\frac{2\sqrt{3}Gm}{r^3}.
\end{eqnarray}
The highest PM order is 2, which here is associated with a typical gravitational field squared $\vec{\Gamma}^2:=G^2m^2/r^4$. Considering only the 2PM terms, we yield an expression for $G_{00}$ that is systematically enhanced compared to $G_{00}\approx R_{00}$ alone: 
\begin{eqnarray}\label{2pm}
G_{00}^\mathrm{2PM}=\frac{2G^2m^2}{r^4}\left(1+2\sqrt{3}\right).
\end{eqnarray}
Here, $(1+2\sqrt{3})\simeq4.4641 \approx4.5$, and recalling the simple assumption of $R=0$ ultimately led to $E=m/6$ at the BH horizon $r=2Gm$, the energy eigenvalue in this case is enhanced roughly by a factor of 4.5 (as a fraction, $9/2$), or
\begin{equation}
\Rightarrow E=\left.\frac{4\pi r^3}{3}\frac{1}{8\pi G}G_{00}^\mathrm{2PM}\right\vert_{r=2Gm}\approx\frac{3}{4}m.
\end{equation}
This is a better estimate than $E=m/6$, reaching roughly three-quarters of the mass-energy relation at the horizon $r=2Gm$. Of course, this is given our choice of coordinates being Eddington-Finkelstein, rather than Schwarzschild's original metric in spherical coordinates. Choosing EF coordinates was an educated choice made out of caution towards the $r=2Gm$ coordinate singularity imposed by the original metric. This coordinate singularity is imposed by the metric comonent $g_{11}=f(r)^{-1}$, which diverges at $r=2Gm$.  However, given that $f(r)$ had been differentiated, removing the integer and focusing more on the second, potential-like term, perhaps our choice of coordinates -- while a free choice -- might come at the price of a systematically reduced accuracy of our expected value. 

Suppose now we revisit the Schwarzschild metric in standard, spherical coordinates \cite{Schwarzschild:1916uq, Schwarzschild:1916ae}:
\begin{equation}\label{schw.sph}
ds^2=-f(r)dt^2+\frac{1}{f(r)}dr^2+r^2d\Omega^2,
\end{equation}
to see if a change in coordinate system alters the accuracy of our approximation. With the metric determinant being $\mathrm{det}(g_{\mu\nu})=-r^4\sin^2\theta$ and $R=0$, and given that all metric elements depend only on $r$ and $\theta$ even still, the Schwarzschild-metric Laplace-Beltrami operator takes on a compact form:
\begin{equation}\label{lbs}
\Delta^\mathrm{LB}_\mathrm{Schw.}=\frac{1}{r^2\sin\theta}\Big[\sin\theta\partial_r\left(r^2f(r)\partial_r\right)+\partial_\theta(\sin\theta\partial_\theta)\Big],
\end{equation}
and using Eq. (\ref{ricci}) with Eq. (\ref{lbs}), one can straightforwardly compute $R_{00}$ to be exactly Eq. (\ref{r00}). Therefore, every derivation step that would follow has already been discussed. Thus, $E\approx0.75m$ is as close as we can reach to our expected BH rest energy value of $E=m$. 

One would be tempted to manually introduce a scaling calibrator to the Kretschmann scalar surrogate, such that $R_\mathrm{eff}=-\lambda\sqrt{K}$. A scaling calibrator was introduced in Ref. \cite{MacKay:2024qxj, MacKay:2025uyg} for the CCB mass-shell model, when it was found that the unscaled surrogate contributed to overestimated GW energy values beyond what is cataloged. Thus, the calibrator was imposed to be $\lambda>1$ to remedy the geometric overshoot imposed by the surrogate, adjusting the contribution to be within the quadrupole regime. In this case, the scaling would be $\lambda<1$ to artificially enhance the inherently optimal geometric overshoot even further. Physically, as Kretschmann scalars encode the Weyl curvature and any present Ricci curvature, the scalar is readily optimized to encode the mass-energy curvature. Imposing an artifical enhancer to the maximal curvature without adding more mass or energy is purely ad hoc, and it should not be pursued.

After all, $E\approx0.75m$ is not a bad estimate. It shows nonetheless that well-defined, relatively simple cases would be underestimated if this variational approach is used liberally. The same principle goes for quantum-mechanical variational method, should one use the method to e.g. solve the Schr\"odinger equation of a free particle. Thus, the original convention of solving the EFEs is reserved for ``simple" systems, e.g. the Schwarzschild and TOV solutions. Perturbative systems with added complexity, such as CCBs generating GWs, or how dark matter (DM) or baryonic matter (BM) may impede GW propagation, can be addressed with this variational methodology, as discussed in Ref. \cite{MacKay:2024qxj}. 

\subsection{Example II: The Linearized Metric Ansatz}

In conventional GR, linearized metric solutions $g_{\mu\nu}\rightarrow\gamma_{\mu\nu}=g^{(0)}_{\mu\nu}+h_{\mu\nu}$ reveal new information regarding first-order, linear perturbations $h_{\mu\nu}$ of a given background metric $g^{(0)}_{\mu\nu}$. Examples include GWs in flat vacuum, whereby $\gamma_{\mu\nu}=\eta_{\mu\nu}+h^\mathrm{GW}_{\mu\nu}$ is the linearized metric consisting of the flat metric $\eta_{\mu\nu}$ and the GW perturbation $h^\mathrm{GW}_{\mu\nu}$; and Regge-Wheeler-gauged, spherical harmonic perturbations of the TOV metric: $\gamma_{\mu\nu}=g^\mathrm{TOV}_{\mu\nu}+h^\mathrm{RW}_{\mu\nu}$, with $h^\mathrm{RW}_{\mu\nu}\propto g^\mathrm{TOV}_{\mu\nu}Y_{lm}(\theta,\phi)$ also containing radial and angular perturbation functions \cite{Regge:1957td, Thorne:1997kt}. In any linearization case, the perturbation is treated as a small deviation relative to the background metric, $|h|\ll1$, which follows that higher-order terms in $h$, $\mathcal{O}(h^2)$, are neglected.

In the context of the Laplace-Beltrami formalism, linearized metric scenarios were first discussed in Ref. \cite{MacKay:2024qxj} when regarding CCBs in the vicinity of e.g. dark matter (DM) backgrounds. DM presents itself as a natural system to introduce as a metric perturbation, as DM is collisionless however gravitational. This follows the logic that DM is not luminous, as it is its namesake, leaving behind a signature along e.g. galaxy clusters as DM halos. Here, DM perturbations are used as a representative example for heuristics. However, the formalism under a linearized metric Ansatz is presented generally for any one's choice and approach of linearization scheme. If we define the linearized inverse metric as $\gamma^{\mu\nu}=g_{(0)}^{\mu\nu}-h^{\mu\nu}$ to motivate $\gamma^{\mu\alpha}\gamma_{\mu\beta}=\delta^\alpha_{~\beta}$ and its square-rooted determinant as $\sqrt{-\gamma}=\sqrt{-g-h}$, the generalized Laplace-Beltrami operator reads as
\begin{equation}
\Delta^\mathrm{LB}_{\gamma}=\frac{1}{\sqrt{-g-h}}\partial_\alpha\left[\sqrt{-g-h}\left(g_{(0)}^{\alpha\beta}\partial_\beta-h^{\alpha\beta}\partial_\beta\right) \right].
\end{equation}
Suppose we assume, however naively, that $\sqrt{-g-h}\simeq\sqrt{-g}$ (i.e., the determinant of the perturbation metric does not contribute much due to its smallness conditions $|h|\ll1$ and $\mathcal{O}(h^2)$), then the determinant of the perturbed metric is screened to be the determinant of the background metric. This allows for the operator to be expressed as a linear but subtractive combination containing the background and perturbed information:
\begin{equation}
\Rightarrow~\Delta^\mathrm{LB}_{\gamma}=\Delta^\mathrm{LB}_g-\frac{1}{\sqrt{-g}}\partial_\alpha\left(\sqrt{-g}h^{\alpha\beta}\partial_\beta \right),
\end{equation}
where $\Delta^\mathrm{LB}_g$ is exactly Eq. (\ref{lbo}), written here as a shorthand. Thus, we may write the explicit operator based on the perturbation metric as $\Delta^\mathrm{LB}_h$. When the linearized operator acts on the metric $\gamma_{\mu\nu}$, we yield
\begin{equation}
\Delta^\mathrm{LB}_{\gamma}\gamma_{\mu\nu}=\Delta^\mathrm{LB}_g g^{(0)}_{\mu\nu}+\Delta^\mathrm{LB}_g h_{\mu\nu}-\Delta^\mathrm{LB}_h g^{(0)}_{\mu\nu}-\Delta^\mathrm{LB}_h h_{\mu\nu},
\end{equation}
where $\Delta^\mathrm{LB}_h h_{\mu\nu}\sim\mathcal{O}(h^2)$ and thus can be neglected. Therefore, the linearized Ricci tensor in this Laplace-Beltrami formalism reads as
\begin{equation}
R_{\mu\nu}=-\frac{1}{2}\left(\Delta^\mathrm{LB}_g g^{(0)}_{\mu\nu}+\Delta^\mathrm{LB}_g h_{\mu\nu}-\Delta^\mathrm{LB}_h g^{(0)}_{\mu\nu} \right)=:R^{(0)}_{\mu\nu}+\delta R^{(0)}_{\mu\nu}-R^{(1)}_{\mu\nu},
\end{equation}
where the orders of $0$ and $1$ in the rightmost linearization denote the perturbation order nested in the Laplace-Beltrami operator. Thus, at ``zeroth order", the linearized Ricci tensor reads rather conventionally: $R_{\mu\nu}\rightarrow R^{(0)}_{\mu\nu}+\delta R^{(0)}_{\mu\nu}$, whereby the Laplace-Beltrami operator is defined only by the background metric $g^{(0)}_{\mu\nu}$.

\subsubsection{Gravitational Waves}

For perturbations from the flat metric and a non-zero GW source term $T^\mathrm{GWS}_{\mu\nu}$, where $\eta_{\mu\nu}$ is the background Minkowski metric and $h_{\mu\nu}$ is a perturbation, $R^{(0)}_{\mu\nu}=0$ -- even in the Laplace-Beltrami formalism --, as $\Delta^\mathrm{LB}_\eta=\Box$ defines the d'Alembert wave operator and $\Box\eta_{\mu\nu}=0$. Therefore, the non-vanishing contribution is the perturbed deviation in the Ricci tensor:
\begin{eqnarray}\nonumber
&\delta R^{(0)}_{\mu\nu}\equiv-\frac{1}{2} \Box h_{\mu\nu}=8\pi G T^\mathrm{GWS}_{\mu\nu}\\\label{GWSeq}
&\Rightarrow~\Box h_{\mu\nu}=-16\pi G T^\mathrm{GWS}_{\mu\nu}.
\end{eqnarray}
The lattermost implication defines the source-on GW equation, under the assumption of a weak-field limit where, effectively, the background metric is flat. However, in the conventional sense, this equation holds for a trace-reversed wave function $\bar{h}_{\mu\nu}=h_{\mu\nu}-\eta_{\mu\nu}h/2$. In the present case, no additional gauges on $h_{\mu\nu}$ were made explicit; one may assign the trace-reversed gauge here as a bookkeeping step: $h_{\mu\nu}\rightarrow\bar{h}_{\mu\nu}$. 

Nonetheless, to solve Eq. (\ref{GWSeq}), either with or without the trace-reversed gauge, we follow convention comfortably by employing Green's function to yield an integral form of the GW function (we inserted here the trace-reversed gauge):
\begin{equation}
\bar{h}_{\mu\nu}(t,\vec{x})=4G\int T^\mathrm{GWS}_{\mu\nu}\,d^3\vec{y},
\end{equation}
where $T^\mathrm{GWS}_{\mu\nu}=T^\mathrm{GWS}_{\mu\nu}(t-D, \vec{y})$, with $D$ being the luminosity distance between the GW source and the observer and $t-D$ being the retarded time for the GW to travel towards the observer. One would then reformulate the integral using Fourier transformation to obtain i.e. the quadrupole moment $Q_{ij}$, where the GW function only evolves over time:
\begin{equation}
\Rightarrow\quad \bar{h}_{ij}(t)=\frac{2G}{D}\ddot{Q}_{ij}.
\end{equation}
It is at this point where one assigns the traceless-transverse (TT) gauge out of necessity, for both the GW function and the quadrupolar source. This yields the very familiar expression for the TT-gauged GW from a source:
\begin{equation}
h^\mathrm{TT}_{ij}(t)=\frac{2G}{D}\ddot{Q}^\mathrm{TT}_{ij}.
\end{equation}

\subsubsection{CCBs in DM Pools}

Now reintroducing the DM example, it is discussed that a CCB may be influenced by the DM pool it may be submerged in. As a heuristics exercise, we consider the ``zeroth order" linearized Ricci tensor, whereby the Laplace-Beltrami operator is defined by the Kerr metric Ansatz (should we view the CCB effectively as a Kerr-like object with a reduced mass measure) under Boyer-Lindquist coordinates, written here in a compact form due to independence in $t$ and $\phi$ \cite{MacKay:2024qxj, MacKay:2025uyg}:
\begin{equation}\label{lbk}
\Delta^\mathrm{LB}_\mathrm{Kerr}=\frac{1}{\Sigma \sin\theta}\left[ \sin\theta \partial_r\left(\Delta  \partial_r\right)+\partial_\theta\left(\sin\theta \partial_\theta\right) \right],
\end{equation}
where $\Sigma=r^2+a^2\cos^2\theta$ with $a=J/m$ being a spin-dependent length-scale, and $\Delta=r^2-2Gm r+a^2$. For the metric Ansatz to describe the CCB, we assert that the mass measure is the reduced mass: $m\rightarrow\mu$. As a representative example, one might write the $\gamma_{00}$ component as a linear combination of the Kerr potential and a DM potential (using the reduced mass as the rotating mass in the Kerr metric):
\begin{equation}\label{trial}
\gamma_{00}=-\left(1-\frac{2G\mu r}{\Sigma} \right)-\Phi_\mathrm{DM}\equiv g^\mathrm{Kerr}_{00}+h^\mathrm{DM}_{00},
\end{equation}
where $\Phi_\mathrm{DM}$ derives from a chosen DM density profile, e.g. a spike $\rho_\mathrm{DM}\approx\rho_0(r/r_0)^{-\chi}$ with $1<\chi<2.5$. For instance, $\Phi_\mathrm{DM}=(4\pi r_0^3/3)G\rho_\mathrm{DM}/r$ to convert the DM density profile as a characteristic mass signature. Therefore, through the linearized decomposition, one recovers the pure Kerr contribution \cite{MacKay:2024qxj, MacKay:2025uyg}:
\begin{eqnarray}
R^{(0)}_{00}=\frac{G^2\mu^2}{2\Sigma^4}\Big(&a^4\left(\frac{3}{2}+\frac{1}{2}\cos(4\theta)\right)\\\nonumber
&+4a^2\left(\cos(2\theta)\left(\frac{a^2}{2}-3r^2\right)-3r^2\right)+4r^4 \Big),
\end{eqnarray}
and the perturbing, DM contribution:
 \begin{equation}
\delta R^{(0)}_{00}=\frac{2\pi G\rho_0}{3\Sigma}\left(\frac{r_0}{r}\right)^{\chi+3}\left(1+\chi\right)\left(a^2(\chi+2)+\chi r^2-2G\mu r (1+\chi)\right).
\end{equation}

In order to assume CCBs as Kerr-like objects, the CCB inclination angle $\iota\in[0,\pi]$ is ``integrated" over when one distributes an equatorial mass-ring of reduced mass measure into a hollow shell. Thus, with the inclination angles serving as effective polar angles $\theta\in[0,\pi]$ we must integrate $R^{(0)}_{00}$ and $\delta R^{(0)}_{00}$ over the full range of $\theta$. This is easily tractable after we recognize that the CCB's spin parameter is comfortably below unity: $a<1$. Thus, after a Taylor expansion over small $a$ and polar-angular integration, we yield 
\numparts
\begin{eqnarray}\label{rook}
&\Rightarrow~R^{(0)}_{00}=\frac{2\pi G^2\mu^2}{r^4}\left(1-5\frac{a^2}{r^2}\right),\\\label{roodm}
&\Rightarrow~\delta R^{(0)}_{00}=\frac{\pi^2 G\rho_0}{3}\left(\frac{r_0}{r}\right)^{\chi+3}\left(1+\chi\right)\\\nonumber&\quad\quad\quad\quad\quad\quad\times\left(2\chi +\frac{a^2}{r^2}(\chi+4)+\frac{2G\mu}{r} (1+\chi)\frac{(a^2-2r^2)}{r^2}\right).
\end{eqnarray} 
\endnumparts
The DM-induced deviation $\delta R^{(0)}_{00}$ presents itself as a PM expansion, whereas the Kerr contribution is of 2PM order. If we discard the 1PM deviation terms, we combine like terms to obtain the total (00)-th Ricci component:
\begin{equation}\label{dmr00}
\Rightarrow~R_{00}=\frac{2\pi G^2\mu}{r^4}\left[{\mu}\left(1-5\frac{a^2}{r^2}\right)-\frac{\rho_0r_0^3}{3}\left(\frac{r_0}{r}\right)^{\chi}\left(1+\chi\right)^2\left(2-\frac{a^2}{r^2}\right)\right].
\end{equation}
It shows in Eq. (\ref{dmr00}) that the DM contribution systematically acts as a damping mechanism to both Newtonian-like and the spin-corrected terms. This supposes that any present DM pool surrounding a CCB supresses the emitted quadrupolar radiation the CCB generates. And this DM-suppression can be, in principle, measured at coalescence; suppose we follow the convention of Kerr spacetime where $R=0$ (out of simplicity, and as a proof of concept), then $T_{00}\approx R_{00}/(8\pi G)$ with total energy $E=T_{00}V$. For the coalescence energy where $r=2GM$, defining $\nu:=\mu/M$ as the symmetric mass ratio,
\begin{equation}
E(t_C) =\frac{\pi}{6}\nu\left[{\mu}\left(1-5\beta_C^2 \right)-\frac{\rho_0r_0^3}{3}\left(\frac{r_0}{2GM}\right)^{\chi}\left(1+\chi\right)^2\left(2-\beta_C^2\right)\right],
\end{equation}
and the difference between the observed radiated energy $E(t_C)$ and the unperturbed energy $\pi\nu\mu(1-5\beta_C^2)/6$ retains the GW energy lost by the DM patch. Since this difference is calculable, the essential DM profile parameters $\rho_0,\,r_0,\,\chi$ can be fitted within respective, physically relevant ranges. Not only this deduces if any DM was present near or around the CCB that sent out GWs, the density profile can be constrained and defined. And more intriguingly, any non-zero residual of the GW energy due to DM dampening may further support microscopic or particle-level mechanisms whereby GWs act as probes on DM \cite{Miller:2025yyx}. 

More intricate details to take into further account is the post-Newtonian (PN) dynamics involved in the inspiral-merger phases of coalescence. It is proposed in Ref. \cite{MacKay:2025uyg} that PN contributions in coalescence reduce the peak energy purely via the orbital dynamics, and the more proper formula obtained in Ref. \cite{MacKay:2025uyg} should be used as the baseline GW energy. This may be subject to a future study, both for heuristic understanding and to provide a benchmark, should future-generation detectors such as the Laser Interferometer Space Antenna (LISA) read signals sourced by supermassive BHs in galactic centers \cite{Amaro-Seoane:2012aqc}.

\subsection{Example III: The Vaidya Metric Ansatz}

The Vaidya metric is a solution to the EFEs for a spherically symmetric and static body that is either emitting or absorbing null dusts (i.e., pools of massless fluid/radiation) \cite{Poisson:2004, Griffiths:2009}. One system that is famously described by this metric is a static BH undergoing quantum-mechanical Hawking radiation \cite{Hawking:1974rv, Hawking:1975vcx}, which we use as a representative example. Under ingoing EF coordinates, such that the BH horizon falls inward overtime due to radiative mass-loss, the Vaidya metric reads \cite{Abdolrahimi:2016emo}:
\begin{equation}
ds^2=-e^{2\psi(z)}\left(1-\frac{2Gm(z)}{r} \right)dv^2+2e^{\psi(z)} dvdr+r^2d\Omega^2,
\end{equation}
where, defining $z=r_S/r$ ($r_S:=2Gm$),
\numparts
\begin{eqnarray}
&\psi(z)\simeq\frac{m_P^2}{m_0^2}\left[g(z)-4\alpha\ln(z)\right],\quad\alpha\approx3.75\times10^{-5},\\
&m(z)\simeq m\left[1+\frac{m_P^2}{m_0^2}h(z) \right],
\end{eqnarray}
\endnumparts
with $m_P\equiv\sqrt{\hbar/G}$ as the Planck mass, $m_0$ as the initial BH mass (which is gauged to be astronomical), and $m$ as the instantaneous BH mass. We can furthermore define the ratio between the squares of the initial BH and Planck masses as an ``initial abundance number'': 
\begin{equation} \label{number}
N_0=\frac{m_0^2}{m_P^2},
\end{equation}
such that the initial BH's surface area $A_0=16\pi G^2m_0^2$ is divided into $N_0$ quantum BH surface areas via $A_Q=16\pi l_P^2$ \cite{Hawking:1971ei, Carr:2005qn}, where $l_P\equiv\sqrt{G\hbar}$ is the Planck length. The functions $g(z)$ and $h(z)$ are polynomials in $z$, given in Ref. \cite{Abdolrahimi:2016emo} as Eq. (85). If we omit higher-order terms of $z$, given the relevant range $z\in(0,1]$, we define
\numparts
\begin{eqnarray}\label{gfull}
&g(z)\simeq\beta{(1-z)(1+z)}\\\nonumber
&\quad\quad\times\left[31k_5-18(\xi-k_6)+6(4f_0-k_3)\frac{(73+49z)}{(1+z)}\right],\\\label{hfull}
&h(z)\simeq\beta{(1-z)(1+z)}\\\nonumber
&\quad\quad\times\left[6\xi+55k_5+18k_6-6(4f_0-k_3)\frac{(23+35z)}{(1+z)}+\frac{144k_4}{(1+z)}\right],
\end{eqnarray}
\endnumparts
where $\beta=({4^{7}\cdot405\pi})^{-1}$, and the parameters $\xi$, $k_{3,4,5,6}$ and $f_0$ depend on the quantum spin of the emitted radiation. It is essential to note that the total power output of Hawking radiation is proposed to consist of 81\% neutrinos, 17\% photons, and 2\% gravitons \cite{Page:1976df}. E.g. for photons (spin-1): $\xi=-1248$, $k_3=81.80$, $k_4=-770.42$, $k_5=65.38$, $k_6=-942.18$, and $f_0=2.4346$ \cite{Abdolrahimi:2016emo, Page:1976ki, Jensen:1990xbb}. Other known values of $\xi$ include $168$ for neutrinos (spin-1/2 particles) and $20352$ for gravitons (spin-2) \cite{Abdolrahimi:2016emo}; the $f_0$ and $k$ parameters are unknown for neutrinos and gravitons. To achieve a spin-independent treatment of Hawking radiation, we examine how the functions $g(z)$ and $h(z)$ impact the metric analysis agnostically. Since $\xi$ is known for all particle types, we consider only the term containing $\xi$ in both functions \cite{MacKay:2025yja}:
\numparts
\begin{eqnarray}\label{g}
&g(z)\approx{-18\xi\beta}(1-z)(1+z),\\\label{h}
&h(z)\approx{6\xi\beta}(1-z)(1+z)=-\frac{1}{3}g(z).
\end{eqnarray}
\endnumparts
The common leading factor, $6\xi\beta$, evaluates numerically as $4.84\times10^{-4}$ for neutrinos, $-3.59\times10^{-4}$ for photons, and $5.86\times10^{-3}$ for gravitons. Expanding out the polynomials and neglecting higher orders of $z$ due to smallness, the metric functions read as
\numparts
\begin{eqnarray}
&\psi(z)\approx-\frac{2}{N_0}\left[9\xi\beta+2\alpha\ln(z)\right],\\
&m(z)\approx m\left(1+\frac{6\xi\beta}{N_0}\right),
\end{eqnarray}
\endnumparts
which modifies the Vaidya metric as follows, in terms of $z$:
\numparts
\begin{eqnarray}\label{vaidya}
&ds^2=-\mathcal{E}(z)\left[1-z\left(1+\frac{6\xi\beta}{N_0}\right) \right]dv^2+2\mathcal{E}(z)^2 dvdr+r^2d\Omega^2,\\
&\mathcal{E}(z):=\left(e^{9\xi\beta}z^{2\alpha}\right)^{N_0/4}=\varepsilon z^{N_0\alpha/2},\quad \varepsilon=\exp\left(\frac{9}{4}\xi\beta N_0\alpha\right).
\end{eqnarray}
\endnumparts
The modification's inclusion of $\xi$ suggests a more explicit treatment of spin-dependent particle types. While the reduced forms of the metric functions are mutual for neutrinos, photons, and gravitons, the analysis on photons should involve the explicit form of the polynomials via Eqs. (\ref{gfull}) and (\ref{hfull}), as all parameters are known.

Being that our new choice of metric Ansatz follows the Eddington-Finkelstein coordinates, we present the Laplace-Beltrami operator in the general form under such coordinates:
\begin{eqnarray}
\Delta^\mathrm{LB}_\mathrm{EF}=\frac{1}{\sqrt{-g}}\Big[&\partial_v\left(\sqrt{-g}g^{vv}\partial_v\right)+\partial_v\left(\sqrt{-g}g^{vr}\partial_r\right)\\\nonumber
&+\partial_r(\sqrt{-g}g^{rv}\partial_v)+\partial_\theta(\sqrt{-g}g^{\theta\theta}\partial_\theta)\\\nonumber
&+\partial_\phi(\sqrt{-g}g^{\phi\phi}\partial_\phi)\Big],
\end{eqnarray}
however the introduction of the variable $z=r_S/r$ slightly complicates the definition of $dv$, $dr$, and their coordinate partial derivatives. One finds that $dz=-(r_S/r^2)dr=-(z^2/r_S)dr$, such that $dr=-(r_S/z^2)dz\Rightarrow\partial_r=-(z^2/r_S)\partial_z$. This redefines $dv=dt+f(r)^{-1}dr$ and $\partial_v$, with $f(r)\rightarrow f(z)=1-z$, as $dv=dt-r_S(z^2(1-z))^{-1}dz$ and $\partial_v=\partial_t-((z^2-z^3)/r_S)\partial_z$. Due to smallness in $z$ through the coordinate's relevant range, we revise $dv=dt-r_Sz^{-2}dz$ and $\partial_v=\partial_t-(z^2/r_S)\partial_z$. As like the previous Schwarzschild example, all components are independent on the coordinate $t$ and $\phi$, allowing us to effectively define $\partial_v\rightarrow -(z^2/r_S)\partial_z$ and drop the term containing $g^{\phi\phi}$:
\begin{eqnarray}
\Delta^\mathrm{LB}_\mathrm{EF}=\frac{1}{\sqrt{-g}}\Big[&\frac{z^2}{r_S^2}\partial_z\left(\sqrt{-g}g^{vv} z^2\partial_z\right)+\frac{z^2}{r_S^2}\partial_z\left(\sqrt{-g}g^{vr} {z^2}\partial_z\right)\\\nonumber
&+\frac{z^2}{r_S^2}\partial_z(\sqrt{-g}g^{rv}z^2\partial_z)+\partial_\theta(\sqrt{-g}g^{\theta\theta}\partial_\theta)\Big],
\end{eqnarray}
where $\sqrt{-g}=\mathcal{E}(z)r^2\sin\theta$ with $r=r_S/z$, and the inverse metric terms $g^{vv},~g^{vr}=g^{rv},~g^{\theta\theta}$ follow from the reciprocal of the coordinate coefficients:
\begin{eqnarray}
\Delta^\mathrm{LB}_\mathrm{Vaidya,~EF}=\frac{z^2}{\mathcal{E}(z)r_S^2\sin\theta}\Big[&-z^2\sin\theta \partial_z\left( \left[1-z\left(1+\frac{6\xi\beta}{N_0}\right) \right]^{-1} \partial_z\right)\\\nonumber
&+2{z^2\sin\theta} \partial_z\left(\frac{1}{\mathcal{E}(z)} \partial_z\right)+\mathcal{E}(z)\partial_\theta(\sin\theta \partial_\theta)\Big].
\end{eqnarray}

Solving for $R_{00}$, such that we can obtain $T_{00}$ under the simple assumption $R=0$, we obtain a cumbersome expression that can be immensely simplified via a Taylor expansion for small $z$:
\begin{equation}
\Rightarrow~R_{00}=\frac{z^2}{8r_S^2}\left(2N_0\alpha-N_0^2\alpha^2 \right) -z^{2-N_0\alpha/2}\frac{N_0\alpha}{2r_S^2\varepsilon}+\mathcal{O}(z^3),
\end{equation}
and with the simplification $R=0$ for the Ricci scalar, we yield the energy density:
\begin{equation}
T_{00}\approx\frac{z^2}{8\pi Gr_S^2}\left[ N_0\alpha\left(\frac{1}{4}-\frac{1}{2\mathcal{E}(z)}\right)-\frac{N_0^2\alpha^2}{8}\right],
\end{equation}
and the energy $E=T_{00}V$ with $V\rightarrow 4\pi (r_S/z)^3/3$. One can notice that as $z\rightarrow0$ ($r\rightarrow\infty$), the energy density diverges due to the $\mathcal{E}(z)^{-1}$ term. Furthermore, the energy eigenvalue $E=T_{00}V$ also diverges due to a large scaling volume. These values would have otherwise be driven to zero due to a typical potential-like drop-off, and it is at this point where one might be cautious to proceed any further. On the other hand, for $z=1$ ($r=r_S$), we have a finite energy density and energy eigenvalue at the BH horizon, recovering $r_S=2Gm$ for the energy eigenvalue:
\numparts
\begin{eqnarray}
&T_{00}\approx\frac{1}{8\pi Gr_S^2}\left[N_0\alpha\left(\frac{1}{4}-\frac{1}{2\varepsilon}\right)-\frac{N_0^2\alpha^2}{8} \right],\\
&E\simeq\frac{1}{3}m\left(N_0\alpha\left(\frac{1}{4}-\frac{1}{2\varepsilon}\right)-\frac{N_0^2\alpha^2}{8} \right).
\end{eqnarray}
\endnumparts

This imposes an inherent issue: the overweighing $N_0^2$ contribution is negative. This suggests that the horizon-level energy of a BH is negative, which is physically impossible and factually incorrect: BHs do \textit{not} harbor negative energy. One might be tempted to ``flip the sign," but this is nonsensical. In this example of metric Ansatz, we must acknowledge that the energy is paradoxically negative under a direct calculation, and we must resort to another alternative to extract a ``true", positive-value calculation. Namely, we consider the energy flux density: $T_{10}$. And given the simplification of $R=0$, $T_{10}\propto R_{10}$, which is purely determined by $g_{10}=g_{rv}=\mathcal{E}(z)^2$ given Eq. (\ref{vaidya}). Solving for $R_{10}$ via the Laplace-Beltrami formalism, we recover another cumbersome expression that we straight-forwardly Taylor expand under a small-$z$ regime:
\begin{equation}
\Rightarrow~R_{10}=\frac{z^2}{2r_S^2}\left(2N_0\alpha-N_0^2\alpha^2 \right) +\frac{z^2\mathcal{E}(z)}{2r_S^2}\left(N_0^2\alpha^2-N_0\alpha \right)+\mathcal{O}(z^3).
\end{equation}
The above polynomial is similar in structure to $R_{00}$, yet there are key details that respect physical intuition. 

Unlike the energy density $T_{00}$, the energy flux density $T_{01}$ vanishes as $z\rightarrow0$. The energy flux $\mathcal{F}=T_{10}A$, however, simplifies nicely given $A=4\pi r^2$ and $r\propto1/z$ (readily considering the overweighing $N_0^2$ terms):
\begin{equation}
\mathcal{F} \approx  \frac{N_0^2\alpha^2}{4G}\left(\mathcal{E}(z)-1 \right)+\mathcal{O}(z).
\end{equation}
Thus, as $z\rightarrow0$ ($r\rightarrow\infty$), the energy flux (proportional to the power emitted $\mathcal{F}\propto\mathcal{P}$) is approximately $-N_0^2\alpha^2/(4 G)$. This is negative, but one could forgive this as this is associated with the \textit{radiated} power as measured from a very-far observer.  At $z=1$, we obtain the surface-level energy flux:
\begin{equation}
\mathcal{F}\equiv\frac{E}{\sigma}\approx \frac{N_0^2\alpha^2\varepsilon}{4G},
\end{equation}
where $\sigma$ is the cross-sectional area. We are reminded that, under the Schwarzschild metric analysis, the energy under $R=0$ is $E=m/6$. Knowing that $N_0=m_0^2/m_P^2$, suppose we extract one mass measure $m_0$ and isolate $m_0/6\rightarrow m/6$ for the energy $E$ (see footnote\footnote{One might be tempted to extract $E=m$ straight-forwardly. However, consistency must be maintained.}). Doing so revises the expression into one for the cross-sectional area:
\begin{equation}
\Rightarrow~ \sigma \approx\frac{2}{3\varepsilon\alpha^2}\frac{\hbar^2}{Gm_0^3}.
\end{equation}

Therefore, not only this entails what we previously found under a different metric Ansatz -- namely, the Schwarzschild metric (and one can claim as a result that $E\approx0.75 mc^2$ in the fuller picture) --, we yield a corresponding cross section $\sigma$ that is infinitesimally small, perhaps unphysically so. Thus, one may view the calculation of $\sigma$ as a byproduct of extracting the desired energy expression. However, if taken seriously as a physical construct, it scales in part by the initial Hawking thermal energy $k_BT_H\sim\hbar/( Gm_0)$ and the characteristic reduced Compton wavelength $\lambda=\hbar/m_0$ that, given $m_0\sim10^{30}$ kg at the bare minimum, is below the Planck length.

In this particular example, we demonstrated that finding the energy eigenvalue is not always as straight-forward as intended. This rests, naturally, in the presentation of the chosen metric Ansatz and how each metric component is defined. In such scenarios where one's calculation of $T_{00}$ leads to unphysical implications, other routes to explore, e.g. a different metric component such as the energy flux density, would provide sensible answers and enable us to extract useful information. On the other hand, one may view this as a limitation to the Laplace-Beltrami formalism, such that in certain circumstances the energy density could not directly yield the energy eigenvalue.

\subsection{Other Effective $T_{\mu\nu}$ Components}

In this Laplace-Beltrami formalism, one can naturally yield other effective components for the energy-momentum tensor via the non-zero components of a given metric Ansatz. Other components of $T_{\mu\nu}$ conventionally include the on-diagonal stress components $T_{ii}$ along the spatial axis $x_i$ ($i=1,2,3$ denotes the 3-spatial coordinates), the 0-th row momentum density $T_{i0}$, the 0-th column energy flux density $T_{i0}$, and the off-diagonal contributions $T_{ij}=T_{ji}$ that typically involve shearing effects. Of course, the free but educated choice of metric Ansatz and coordinate-system affects the triviality of certain components. Suppose we have arbitrary metric Ans\"atze with cross terms:
\numparts
\begin{eqnarray}
&\mathrm{Cartesian+Shear:}\quad ds^2=ds_\mathrm{Cart.}^2+g_{ij}dx_idx_j+g_{ji}dx_jdx_i,\\
&\mathrm{Spherical+Spin:}\quad ds^2=ds_\mathrm{Sph.}^2+g_{03}dtd\phi+g_{30}d\phi dt.
\end{eqnarray}
\endnumparts

From this list of select coordinates, and the example Ans\"atze we discussed, one can be cetain that the energy density $T_{00}$ is non-zero via the 00-th time-like component $g_{00}$. Thus, one can expect the stress values $T_{ii}$ to be derived from any non-zero $g_{ii}$ contribution, e.g. under Cartesian and spherical coordinates and their respective expansions. That is, with the exception being the $g_{11}$ component under Eddington-Finkelstein coordinates (it is zero), resulting to a radial-direction stress and its corresponding pressure being trival. For the Eddington-Finkelstein-coordinated Schwarzschild metric, this parallels with the Oppenheimer-Snyder picture of BHs as being dust-like \cite{Oppenheimer:1939ue}, however one could define an effectively non-zero radial-direction stress under e.g. the spherical-coordinated Schwarzschild metric. This is given a non-zero $g_{11}$ contribution via $g_{11}=f(r)^{-1}$. 

Suppose we pursue this, purely for heuristics. We recall the spherical Schwarzschild metric via Eq. (\ref{schw.sph}); under the respective Laplace-Beltrami operator given as Eq. (\ref{lbs}), we yield the $R_{11}$ expression that relates to the radial stress $T_{11}=\sigma_{rr}$:
\begin{equation}\label{rs11}
R_{11}=-\frac{2G^2m^2}{r^4}\left(1-\frac{2Gm}{r} \right)^{-2}.
\end{equation}
One of the notorious nuances of Schwarzschild's original, spherical-coordinated metric solution was the coordinate singularity at $r=2Gm$, which is caused by the produced asymptote within $1/f(r)$. Because of the coordinate singularity, the effective pressure determined by $R_{11}\propto\Delta^\mathrm{LB}_\mathrm{Schw.}g_{11}$ diverges to rather unphysical proportions. If one chooses a regulation scheme such that $R_{11}$ is Taylor expanded under small $r/(2Gm)=:x$, and one drives (albeit naively) $x\rightarrow1$, we yield:
\begin{equation}
\Rightarrow\quad R_{11}=-\frac{1}{8G^2m^2x^2}-\frac{1}{4G^2m^2x}-\frac{3}{8G^2m^2}-\frac{x}{2G^2m^2}+\mathcal{O}(x^2) .
\end{equation}
Should one have allowed additional orders of $x$ into the expansion, one can see that driving $x\rightarrow1$ indeed leads to a divergence -- not one via an asymptote in the function, but rather via a non-convergent summation of terms. For completeness, should one drive $x\rightarrow0$ towards the geometric singularity, there exists only one convergent term hidden by the vanishing and infinite divergences: $-3/(8G^2m^2)$. 

On the other hand, returning to $x\rightarrow1$, if we allow a cutoff at e.g. $\mathcal{O}(x^2)$, one obtains a convergent value that is nonetheless reflective on the choice of cutoff order:
\begin{equation}
\lim_{x\rightarrow1}R_{11}=-\frac{5}{4G^2m^2} +\mathcal{O}(x^2).
\end{equation}
This defines the radial stress $\sigma_{rr}$ to be negative (i.e., pointing radially inward), once we include the Kretschmann scalar contribution: 
\begin{equation}
\sigma_{rr}=-\frac{1}{32\pi G^3m^2}\left[5+4\sqrt{3}\right]\approx-\frac{3}{8\pi G^3m^2}.
\end{equation}
This shows that the pressure, defined by the average stress via the generalized trace formula:
\begin{equation}\label{press}
P=-\frac{1}{\mathrm{n}}\sum_{i=1}^\mathrm{n}\sigma_{ii},
\end{equation}
in which typically $\mathrm{n}=3$ in 3-space, is non-trivial. Furthermore, this stress component is inward-pointing, towards the geometric singularity, and it is dependent on $1/m^2$. That is, the largeness of the BH mass gauges the smallness of the resulting pressure, effectively recovering dust-like null pressure for astrophysical BHs (i.e., the Oppenheimer-Snyder picture). 

One can make the logical intuition that, given the stress is inward-pointing, the respective pressure is associated with either: (i) the trapped surface topology introduced by Roger Penrose \cite{Penrose:1964wq}, or (ii) the cause behind the gradual inward drag of the BH horizon due to Hawking radiation and radiative mass-loss. Through the second intuition, we can naturally pursue the ideal gas law from this pressure, assuming briefly that the Schwarzschild metric describes a BH undergoing Hawking radiation (see footnote\footnote{This assumption was made in Ref. \cite{MacKay:2025yja}, however under ingoing EF coordinates.}). 

We recognize, under the spherical Schwarzschild metric, the other stress components $\sigma_{\theta\theta},~\sigma_{\phi\phi}$ via $g_{22}=r^2$ and $g_{33}=r^2\sin^2\theta$, respectively, determine a polar- and azimuthal-angular pressure. One can find under the Laplace-Beltrami formalism that the corresponding effective $G_{22}$ and $G_{33}$ calculations are of 1PM order, compared to the 2PM order of Eq. (\ref{rs11}). Thus, we discard the 1PM order terms, which only leaves the effecitve 2PM $G_{11}$ component and its associating radial pressure $\sigma_{rr}$ after Taylor expansion and setting $x\rightarrow1$. From which, we effectively define the total pressure via Eq. (\ref{press}), with $\mathrm{n}=3$ to cover all 3 spatial dimensions:
\begin{equation}
P\approx \frac{1}{8\pi G^3m^2}.
\end{equation}
Via $P=nk_BT$ ($n:=N/V$ is the number density and $k_BT$ is the system's thermal energy) and using the Hawking thermal energy $k_BT_H=\hbar/(8\pi Gm)$, we calculate the associating number density for the Hawking radiation:
\begin{equation}
n\approx \frac{1}{\hbar G^2m}=\frac{2}{l_P^2 r_S}.
\end{equation}
redefining $r_S=2Gm$ and $l_P=\sqrt{\hbar G}$. From the definition $n=N/V$ and the effective BH volume $V=4\pi r_S^3/3$, the total number of Hawking radiation particles (stored along the BH surface to be released) follows the expression:
\begin{equation}
N\approx \frac{8\pi r_S^2}{3l_P^2}=\frac{2A}{3l_P^2}.
\end{equation}

We remind ourselves that the Bekenstein-Hawking formula for BH entropy is $S=k_BA/(4l_P^2)$, where $A$ is the BH surface area $4\pi r_S^2$. The ratio $A/(4l_P^2)$ yields $\pi r_S^2/l_P^2$, which our calculation for $N$ scales by a factor of $8/3=2.667$. Therefore, one can make the claim that an enhanced (more than double) Bekenstein-Hawking entropy for BHs is recovered under the demonstrated regulated treatment in the Laplace-Beltrami formalism. This perserves the statistical-mechanical convention that the entropy roughly scales by the number of particles in the system: $S\sim k_BN$ (conventionally by the natural logarithm of the microstate multiplicity: $S=k_B\ln\Omega$, whereby $\ln\Omega\sim N$). One has to recognize that this scaling enhancement is a byproduct of using the variational and regulation techniques used to derive this relation.

\section{First-Order Laplace-Beltrami}

Via Eq. (\ref{1stord}), the EFEs take on a vector gradient form that is strongly background dependent, presenting itself as a first-order differential equation while maintaining its tensor-2 ranking. While the EFEs are inherently second-order differential equations, it is of interest to see if this differential-order promotion is persistent even in the first-order Laplace-Beltrami formalism. In other words, it is of interest to define the 4-vector as a covarient gradient of a scalar potential, much like its Euclidian counterpart: $\vec{V}=-\vec{\nabla}\varphi$. To do so, we utilize Eq. (\ref{veceq}) and $R=\nabla^\mu V_\mu$. 

We contract the Ricci coupling in Eq. (\ref{veceq}) with the metric $g^{\sigma\nu}$ to define the respective Ricci scalar:
\begin{equation}
g^{\sigma\nu}\Delta^\mathrm{LB}V_\nu=RV^\sigma,
\end{equation}
to which we multiply with the vector $V_\sigma$. This is to be contracted by the metric on the left hand side, and to introduce the inner product $||V||^2=V^\sigma V_\sigma$ on the right hand side. The aim is to isolate the Ricci scalar, so that we can equate it to $R=\nabla^\mu V_\mu$ derived in the Bianchi identity sanity check:
\begin{equation}
R=\frac{1}{||V||^2}V^\nu\Delta^\mathrm{LB}V_\nu\equiv\nabla^\nu V_\nu.
\end{equation}
As we are in a metric-independent treatment, we can freely define $\Delta^\mathrm{LB}=\nabla_\nu\nabla^\nu$, from which we define
\begin{equation}
\frac{1}{||V||^2}V^\nu\nabla_\nu\nabla^\nu V_\nu = \nabla^\nu V_\nu.
\end{equation}
This cancels out the common $\nabla^\nu V_\nu$, leaving an open operator $\nabla_\nu$ to define a scalar, e.g. unity. Mathematically, defining any function or quantity with an open operator is forbidden. I.e. an n-dimensional vector must be defined by the n-dimensional gradient of a scalar function $\varphi$ to appoint direction and dimensionality. Therefore, we introduce $\varphi$ to close the open operator, which defines furthermore:
\begin{equation}
V^\nu\nabla_\nu\varphi= V^\nu V_\nu.
\end{equation}
If one removes the common $V^\nu$, we define the 4-vector as a 4-gradient acting on a scalar function: $V_\mu=\nabla_\mu\varphi$, and $V^\mu=\nabla^\mu\varphi$. Being that $V^\mu=(V_0,~\vec{V})$, we define the 4-vector components as follows, under any spacetime:
\begin{equation}
V_0=\frac{1}{\sqrt{-g}}\partial_t \left(\sqrt{-g}g^{0\nu}\varphi\right), \quad \vec{V}=\frac{1}{\sqrt{-g}}\partial_i \left(\sqrt{-g}g^{i\nu}\varphi\right).
\end{equation}
E.g., for a flat spacetime where $g_{\mu\nu}=\eta_{\mu\nu}$ and $\sqrt{-\eta}=1$, we yield the expressions (in the $\eta=\mathrm{diag}(-,+,+,+)$ signature):
\begin{equation}
V_0=-\partial_t \varphi, \quad \vec{V}=\vec{\nabla}\varphi.
\end{equation}
To recover the minus sign for the 3-gradient, one would use the alternative signature $\eta=\mathrm{diag}(+,-,-,-)$, which is a convention not typically used in GR yet widely used in particle physics and quantum field theory. One must keep in mind that using the alternative signature must be consistent for the other order terms, i.e. second and zeroth. 

Providing our obtained definition for the 4-vector as a 4-gradient on a scalar function, we modify Eq. (\ref{1stord}) such that it \textit{becomes} a second-order form:
\begin{equation}
G_{\mu\nu}^{(1)}=\frac{1}{2}\nabla_{(\mu}\nabla_{\nu)}\varphi-\frac{1}{2}g_{\mu\nu}\left(\nabla^\mu \nabla_\mu\varphi\right).
\end{equation}
The first term, $\nabla_{(\mu}\nabla_{\nu)}\varphi$, defines the Hessian tensor $\mathbf{H}_{\mu\nu}(\varphi)$, which in the first-order Laplace-Beltrami formalism acts as an effective Ricci tensor. In the second term, the Laplace-Beltrami operator naturally emerges: $g_{\mu\nu}\left(\nabla^\mu \nabla_\mu\varphi\right)=g_{\mu\nu}\Delta^\mathrm{LB}\varphi$. For the right hand side of the EFEs, $8\pi GT^{(1)}_{\mu\nu}$, the first-order energy-momentum tensor must be coupled to the scalar function $\varphi$ for mathematical completeness, i.e. $T^{(1)}_{\mu\nu}= T_{\mu\nu}\varphi$ rather simply. Therefore, the EFEs in the first-order Laplace-Beltrami formalism is, essentially, a second-order differential equation to a scalar field function:
\begin{equation}\label{hess}
\frac{1}{2}\left( \mathbf{H}_{\mu\nu}(\varphi)-g_{\mu\nu}\Delta^\mathrm{LB}\varphi\right)=8\pi GT_{\mu\nu}\varphi.
\end{equation}
We can impose further analytical control on this equation by contracting Eq. (\ref{hess}) with the background metric $g^{\mu\nu}$. Thus, under any choice of metric Ansatz, contracting Eq. (\ref{hess}) with $g^{\mu\nu}$ leads to the following inhomogeneous equation:
\begin{equation}\label{scaleq1}
\Rightarrow -\Delta^\mathrm{LB}\varphi=8\pi GT_{\mu}^{~\mu}\varphi,
\end{equation}
which takes the form of a Klein-Gordon equation on the given curved spacetime. Much like Eq. (\ref{veceq}), this equation is background-dependent; however, this equation is source-dependent rather than being curvature-coupled. To remedy this, one would take the extra step in defining $8\pi GT_\mu^{~\mu}=G_\mu^{~\mu}$, which is conventionally the negated Ricci scalar $-R$. To motivate a scalar field equation of motion similar in form to Eq. (\ref{veceq}), we yield:
\begin{equation}\label{scaleq2}
\Delta^\mathrm{LB}\varphi=R\varphi.
\end{equation}
As like the second-order Laplace-Beltrami formalism, both $\Delta^\mathrm{LB}$ and $R$ are defined by one's choice of metric Ansatz, with the operator keeping its time-differential terms for scalar wave propagation and $R$ either defined by the Laplace-Beltrami trace or effectively via the Kretschmann scalar surrogate.

\subsection{Example I: The Flat (Minkowski) Metric Background}

For the flat background, we will consider a pure vacuum where $R=0$ conventionally and an impure yet flat vacuum with a non-zero source $R=-8\pi GT^\mu_{~\mu}$. Firstly, for a flat vacuum, Eq. (\ref{scaleq2}) reduces perfectly to a d'Alembert wave equation for a scalar field:
\begin{equation}\label{kgeq}
 \Box\varphi=0.
\end{equation}
This equation recovers the massless Klein-Gordon equation for spin-0 fields, just as how Eq. (\ref{s1f}) takes on a massless Klein-Gordon equation for spin-1 fields. The solution to this equation straight-forwardly resembles Wentzel-Kramers-Brillouin (WKB) plane waves \cite{Birrell:1982ix}:
\begin{equation}
\varphi(x)\sim \exp(-ik_\mu x^\mu),
\end{equation}
where $k^\mu=(\omega, \vec{k})$ is a wave number 4-vector with $x^\mu=(t,\vec{x})$ on a flat spacetime.

For the impure yet flat vacuum with a non-zero source, suppose the energy-momentum trace is screened, e.g. $T^\mu_{~\mu}=\epsilon^2/(8\pi G)$. In this way, the spacetime background still follows a Minkowski metric Ansatz. The right hand side of Eq. (\ref{kgeq}) is here non-zero:
\begin{equation}
\Rightarrow -\Box\varphi=\epsilon^2\varphi.
\end{equation}
Purely for heuristics, suppose we define the gauge $\epsilon=\mathrm{m}/\hbar$, where $\mathrm{m}$ is the particle rest mass and $\hbar$ is the reduced Planck constant. This choice of gauge recovers the massive Klein-Gordon equation:
\begin{equation}
-\hbar^2\Box\varphi=\mathrm{m}^2\varphi.
\end{equation}

\subsection{Example II: The Schwarzschild Background}

Under the Schwarzschild metric Ansatz for the background, Eq. (\ref{scaleq2}) reads as
\begin{equation}
\Delta^\mathrm{LB}_\mathrm{Schw.}\varphi=R\varphi,
\end{equation}
where the time-independent Laplace-Beltrami operator under the Schwarzschild metric Ansatz is identical under spherical coordinates and EF coordinates. For our scalar function solution, we must utilize the time-component in e.g. the spherical coordinated operator to yield wave propagation. As previously suggested, the Ricci scalar $R$ can either be defined by: (i) the Laplace-Beltrami trace, or (ii) effectively via the Kretschmann scalar surrogate. As we know that certain Laplace-Beltrami-derived Ricci scalars are divergent, such as the spherical-coordinated Schwarzschild case, we consider the second option, which leads to the following partial differential equation to solve:
\begin{eqnarray} \label{scalschw}
\frac{1}{r^2\sin\theta}\Big[-\frac{r^2\sin\theta}{f(r)}\partial^2_t+\sin\theta\partial_r\left(r^2f(r)\partial_r\right)+&\partial_\theta(\sin\theta\partial_\theta)\Big]\varphi(r,t,\theta)\\\nonumber
&=-\frac{4Gm\sqrt{3}}{r^3}\varphi(r,t,\theta).
\end{eqnarray} 
We can furthermore gauge the polar angle $\theta$ to be e.g. at the equatorial plane $\pi/2$; this gives us the time and radial contributions with $\sin(\pi/2)=1$, with the scalar function now dependent on $r$ and $t$, such that $\varphi(r,t)\equiv B(r)T(t)$ via separation of variables. Recalling that $f(r)=1-2Gm/r$, we find the stationary solution for $B(r)$, independent on $t$, defined in terms of the hypergeometric ${}_2F_1$ functions:
\begin{eqnarray}
B(r)\simeq&3.293a_0\left(\frac{Gm}{r}\right)^{1.861}{}_2F_1\left(-1.861,\,-0.8612;-\,2.722;\,\frac{r}{2Gm}\right)\\\nonumber
&+0.2495a_1\left(\frac{r}{Gm}\right)^{1.861}{}_2F_1\left(1.861,\, 2.86121;\,4.722;\,\frac{r}{2Gm}\right).
\end{eqnarray}
Here, $a_0$ and $a_1$ are integration constants. Introducing the non-trivial time contribution yields a characteristic time-dependent propagator: $\varphi(r,t)\sim B(r)\exp(i\omega t)$.

One recognizes that the hypergeometric functions, regardless of the value of $t$ or any non-variable arguments, exist within $r/(2Gm)\in(0,1)$. Therefore, the provided solution is an interior solution, which becomes wave-like overtime and diverges at the asymptotes $r=0$ and $r=2Gm$. For exterior solutions, we must invoke the weak Schwarzschild limit, such that $r\rightarrow\infty$ asymptotically, whereby $f(r)\rightarrow1$. This revises Eq. (\ref{scalschw}) as
\begin{equation}
\Rightarrow\quad\Big[-\partial^2_t+\frac{1}{r^2}\partial_r\left(r^2\partial_r\right)\Big]\varphi(r,t)=-\frac{4Gm\sqrt{3}}{r^3}\varphi(r,t),
\end{equation} 
from which we yield the real, stationary solution as a Bessel funtion of the first kind:
\begin{equation}\label{weaks}
B(r)=2a_0\sqrt[4]{3}\sqrt{\frac{Gm}{r}}J_1\left(4\sqrt[4]{3}\sqrt{\frac{Gm}{r}} \right).
\end{equation}
The non-zero time contribution once again recovers the time-dependent propagator. Given the relevant range is $r\in[2Gm,\infty)$, one finds that $B(r)$ is finite at $r=2Gm$ and exponentially decays to zero as $r\rightarrow\infty$. In other words, we obtain a wave that decays into a zero flat-line as our radial coordinate extends to asymptotic flatness. This does not recover the pure vacuum solution to Eq. (\ref{kgeq}).

\subsection{Example III: A Cartesian Schwarzschild Background}

Eq. (\ref{weaks}) does not recover the flat-metric planar wave solution as $r\rightarrow\infty$, i.e. on an ultra-weak Schwarzschild, Minkowskian-like spacetime. This would motivate a new analysis under a change in coordinate system, as one would recall that the flat metric is canonically Cartesian. Thus, instead of spherical (or EF) coordinates that yield $\sqrt{-g}=r^2\sin\theta$, which is the culprit behind yielding both the hypergeometric and Bessel function solutions, we consider isotropic Cartesian coordinates \cite{Choquet-Bruhat:2014}:
\begin{equation}
ds^2_\mathrm{Schw.iso}=-\left(\frac{2\rho-Gm}{2\rho+Gm}\right)^2dt^2+\left(1+\frac{Gm}{2\rho}\right)^4(dx^2+dy^2+dz^2).
\end{equation}
Here, the original radial coordinate is transformed into $r\rightarrow\rho(1+Gm/(2\rho))^2$, where $\rho$ is the new radial coordinate. This new radial variable can decompose into the Cartesian components via the sphere radius expression $\rho^2=x^2+y^2+z^2$. The inverse metric $g^{\mu\nu}$ is defined by the reciprocal terms of the metric elements, and the determinant is a polynomial in $\rho^{-1}$: $\mathrm{det}(g_{\mu\nu})=-2+G^2m^2/\rho^2-G^4m^4/\rho^4:=g(\rho^{-1})$. Therefore, when writing the Laplace-Beltrami operator, we essentially have an augmented d'Alembert wave operator whereby $\partial_i\rightarrow\partial_\rho$:
\begin{eqnarray}
\Delta^\mathrm{LB}_\mathrm{Schw.iso}=&-\left(\frac{2\rho+Gm}{2\rho-Gm}\right)^2\partial^2_t\\\nonumber
&+\frac{1}{\sqrt{-g(\rho^{-1})}}\partial_\rho\left( \sqrt{-g(\rho^{-1})} \left(1+\frac{Gm}{2\rho} \right)^{-4}\partial_\rho\right) .
\end{eqnarray}
We note that the physical range of the new radial coordinate is within $\rho\in(Gm/2,\infty)$, where $\rho=Gm/2$ is the transformed horizon scale under isotropic coordinates; this recovers $r=2Gm$ under the old radial coordinate. For $\rho\rightarrow\infty$, the isotropic-coordinated Laplace-Beltrami operator asymptotically becomes the d'Alembert wave operator. Considering briefly the stationary solution with $t=0$ and no time-differential contribution, the right-hand side of Eq. (\ref{scalschw}) needs to be rewritten in terms of the new radial coordinate:
\begin{eqnarray}\label{scalschw2}
\frac{1}{\sqrt{-g(\rho^{-1})}}\partial_\rho\Bigg( \sqrt{-g(\rho^{-1})} &\left(1+\frac{Gm}{2\rho} \right)^{-4}\partial_\rho\Bigg)B(\rho)\\\nonumber
&=-\frac{4Gm\sqrt{3}}{\rho^3}\left(1+\frac{Gm}{2\rho}\right)^{-6}B(\rho).
\end{eqnarray} 
Eq. (\ref{scalschw2}) in itself is a cumbersome ordinary differential equation to solve explicitly. Taming this requires e.g. regulation schemes under the strong and weak Schwarzschild limits. Respectively, $\rho$ is treated asymptotically as small and large; under the weak Schwarzschild (large-$\rho$) limit,
\begin{eqnarray}
\Rightarrow\quad\partial^2_\rho B_w(\rho)=-\frac{4Gm\sqrt{3}}{\rho^3}B_w(\rho),
\end{eqnarray}
which yields a real solution in terms of a Bessel function of the first kind:
\begin{equation}
B_w(\rho)=\frac{a_0}{2\sqrt[4]{3}}\sqrt{\frac{\rho}{Gm}}J_1\left(4\sqrt[4]{3}\sqrt{\frac{Gm}{\rho}} \right).
\end{equation}
One finds that the obtained stationary solution is a position-dependent amplitude modifier for the time-dependent wave propagator, once the time-differential contribution is recovered and solved for. E.g., $B_w(\rho)$ plateaus at a certain, constant value as $\rho\rightarrow\infty$, recovering the planar wave profile for an asymptotically flat background. On the other hand, under the strong Schwarzschild (small-$\rho$) limit,
\begin{eqnarray}
\Rightarrow\quad\partial_\rho\left( \rho^2 \partial_\rho\right)B_s(\rho)=-16\sqrt{3}\frac{\rho}{Gm}B_s(\rho),
\end{eqnarray} 
which also yields a real solution in terms of the Bessel function: 
\begin{equation}
B_s(\rho)=\frac{a_0}{4\sqrt[4]{3}}\sqrt{\frac{Gm}{\rho}}J_1\left(8\sqrt[4]{3}\sqrt{\frac{\rho}{Gm}} \right).
\end{equation}
We note that the stationary solution itself exhibits a wave-like behavior in the range $\rho\in(Gm/2,\infty)$, where the function tapers off towards zero as $\rho>1$ while maintaining its profile. 

Provided these weak and strong limit solutions, respectively $B_w(\rho)$ and $B_s(\rho)$, one becomes motivated to stitch these limits together and form a continuous function over all $\rho$. If one does this, this could be done either via: (i) a piecewise stitching, or (ii) a linear combination between the limit solutions, provided that each respective function has tapering limits and convergences at specific amplitudes. Suppose we pursue the second option, a choice made to avoid any nonsensical ``kinks" in the continuous function; the time-dependent scalar function upon the isotropic Schwarzschild background reads as
\begin{eqnarray}\label{schwsol}
&\varphi(\rho,t)=\Big(B_s(\rho)+B_w(\rho) \Big)\exp\left(i\omega t\right)\equiv B(\rho)\exp\left(i\omega t \right),\\\nonumber
&\mathrm{where}\quad\rho\in\left(\frac{Gm}{2},\infty\right).
\end{eqnarray}  

\subsection{Discussion}

Obtaining the relevant equations of motion upon any curved background, e.g. Eq. (\ref{veceq}) for 4-vectors and Eq. (\ref{scaleq2}) for scalar functions, are reminiscent to previous attempts to approach quantum field theory in curved backgrounds (for related literature, see e.g. \cite{Bunch:1979uk, Bekenstein:1981xe, Alsing:2000ji, Li:2024ltx}). Thus, using these background-dependent wave equations, one is encouraged to define e.g. Feynman rules from these equations, such that one may use second quantization to Fourier-transform 4-spatial partial derivatives into 4-momentum variables, and thus any 4-spatial variables into 4-momentum partial derivatives. And one may do so after designating the relevant background the scalar and vector fields reside on.

We note the work done in Ref. \cite{Li:2024ltx}, whereby the authors claimed to have derived curved spacetime modifications to Feynman diagrams for spin-0 scalar and spin-1/2 spinor particles -- spin-1 vector particles were never discussed. However, the analysis was done under the simplified scenario of local Minkowski coordinates and a weak Schwarzschild background ($r\gg 2Gm$). It is also worth noting that the authors claimed to have found a generalized form of the Klein-Gordon equation for any background as their Eq. (2.6) (i.e., a form similar to Eq. (\ref{scaleq2}) with the right hand side instead containing the particle rest mass $\mathrm{m}^2\varphi$ with $\hbar=1$). However, the authors' ad hoc Klein-Gordon generalization (i.e., $\Box\rightarrow\Delta^\mathrm{LB}$ without regard of further curvature implications) is naive, even under the case of a weakly-gauged Schwarzschild background. On this basis, among further critiques formulated in Ref. \cite{MacKay:2025dbu}, we reject the claims made in \cite{Li:2024ltx}.

\section{Zeroth-Order Laplace-Beltrami}

Via Eq. (\ref{0thord}), the EFEs take on a zeroth-order form, i.e. the Ricci tensor is written in terms of an auxiliary tensor $A_{\mu\nu}$ that is of zeroth-differential order. As a term within a composition containing inherently second-differential order terms: Eq. (\ref{2ndord}) via the Laplace-Beltrami operator and Eq. (\ref{1stord}) via the covariant divergence of a 4-vector (which itself is a 4-gradient of a scalar function), one might be tempted to follow the logic that the zeroth-order term must somehow be of second-differential order for consistency. At first glance, this appears to impose an incompatibility, as $A_{\mu\nu}$ was introduced to be arbitrary, similarly to the integration constant $+C$. 

This apparent issue can be resolved by recognizing that the auxiliary tensor serves as an effective gauge/screening mechanism that can constrain e.g. propagating degrees of freedom. This constraining role becomes more transparent under a transformation to a symbolic representation, such as in Fourier space. Should e.g. $A_{\mu\nu}=\varpi^2 g_{\mu\nu}/2$, where $\varpi$ is a constraint parameter with its square containing curvature units, then the Fourier transformation $A_{\mu\nu}\rightarrow\widetilde{A}_{\mu\nu}$ implies the corresponding transformation of the leading higher-order terms. In local coordinates, $\partial_\alpha\rightarrow -i k_\alpha$, while in general $g_{\mu\nu}\rightarrow\widetilde{g}_{\mu\nu}$; recall that $k^\alpha=(\omega,\vec{k})$ denotes the local 4-wave vector. 

If we heuristically treat the ``Fourier spacetime" as background-dependent, then to leading order in a local Fourier representation the physical spacetime covariant derivatives may be represented non-locally by a background-dependent 4-momentum vector $\mathbf{k}^\alpha$, i.e. $\nabla_\alpha\rightarrow -i\mathbf{k}^\alpha$ non-locally. Under this correspondence, the Laplace-Beltrami formalism for the Fourier-transformed Ricci tensor takes the following form:
\begin{equation}
\widetilde{R}_{\mu\nu}\simeq\frac{1}{2}\left(\mathbf{k}_\alpha \mathbf{k}^\alpha -\varpi^2 \right)\widetilde{g}_{\mu\nu}-\frac{i}{2}\mathbf{k}_{(\mu}\widetilde{V}_{\nu)},
\end{equation} 
where the real component contains the mass-shell condition $\mathbf{k}_\alpha \mathbf{k}^\alpha -\varpi^2$ governing dispersion. In this sense, $\varpi$ may be interpreted as analogous to a physical mass scale or a gauge-dependent infrared regulator. 

However, we emphasize that this Fourier spacetime description is a representative device used to elucidate the role of the auxiliary tensor in physical spacetime. Accordingly, $A_{\mu\nu}$ encodes gauge and background information that is otherwise left implicit in the first- and second-order decompositions previously discussed. This is while leaving the leading second-order Laplace-Beltrami character of the Ricci tensor unchanged.

\section{Concluding Statements}\label{sect:concl}

In this work, we explored in detail the Laplace-Beltrami formalism of the EFEs, whereby one defines at face-value the Ricci tensor in terms of the Laplace-Beltrami partial differential operator. One therefore uses a variational methodology to solve the EFEs under a choice of metric Ansatz and coordinate system, analogously to solving impossible Hamiltonians that describe pertrubative quantum-mechanical systems. However, the focus of this second-order analysis had been on well-known ``simple" systems such as static spherical bodies, and how the Laplace-Beltrami-formulated results compare to conventional understanding. It is therefore encouraged for future studies to use the formalism on perturbative extensions to ``simple" astrophysical systems, e.g. a rotating \cite{Gamow:1946hyv, Godel:1949ga, Su:2009fu} or a bulk-viscous \cite{Brevik:2014cxa, Choudhuri:2015, Srivastava:2018gye, Hu:2020xus, Arora:2021tuh, Singh:2023evc} FLRW universe, and superfluid models of mirrored BH interiors \cite{Hayden:2007cs, Manikandan:2018urq}. It is of course natural to propose further analyses on additional perturbative astrophysical systems, either continuing into GWs and various environmental factors or veering into alternate directions.

Provided the first-order analysis was able to define (i) curvature-coupled wave equations for 4-vectors (Eq. [\ref{veceq}]) and scalar functions (Eq. [\ref{scaleq2}]), and (ii) a rendering of the EFEs in the structure of a scalar wave equation via Eq. (\ref{hess}),  it is of great interest to pursue each of these routes individually. In the case of the curvature-coupled wave equations, analyses of quantum- or effective-field theory on curved spacetimes can be approached using the acquired formulae:  the 4-vector equation for spin-1 particles, and the scalar equation for spin-0 particles. In this endeavor, one may use the second-order Laplace-Beltrami definition of the Ricci tensor for spin-2 particles, more prominently gravitons. As for the scalar-function rendering of the EFEs, it is an intellectual curiosity to explore GR in the context of a manifold dominated by a scalar field, e.g. the cosmological epoch of the universe when the Higgs field is dominant \cite{Calmet:2017hja, Steinwachs:2019hdr}, or direct analysis of scalar-field DM candidates  \cite{Sahni:1999qe, Matos:2000ng, Matos:2000ss, Alcubierre:2001ea, Bernal:2006it}.


\section*{Statement Declarations}

\subsection*{Conflict of Interest}
The author declares no conflicts of interest.

\subsection*{Data Access Statement}
As a theoretical study, this work generates no original data. 

\subsection*{Ethics Statement}
No ethical issues arise, as no test subjects are involved. This paper adheres to academic integrity.

\subsection*{Funding Statement}
This work received no funding.







\section*{References}

\begin{thebibliography}{99}


\bibitem{Einstein:1915ca}
Einstein~A 1915 \textit{
Sitzungsber. Preuss. Akad. Wiss. Berlin (Math. Phys. )} \textbf{1915}, 844

\bibitem{Einstein:1915bz}
Einstein~A 1915 \textit{
Sitzungsber. Preuss. Akad. Wiss. Berlin (Math. Phys. )} \textbf{1915}, 831

\bibitem{Dyson:1920cwa}
Dyson~F~W, Eddington~A~S and Davidson~C 1920 \textit{
Phil. Trans. Roy. Soc. Lond. A} \textbf{220}, 291


\bibitem{Hubble:1926yw}
Hubble~E~P 1926 \textit{
Astrophys. J.} \textbf{64}, 321

\bibitem{Hubble:1929ig}
Hubble~E 1929 \textit{
Proc. Nat. Acad. Sci.} \textbf{15}, 168

\bibitem{Schwarzschild:1916uq}
Schwarzschild~K 1916 \textit{
Sitzungsber. Preuss. Akad. Wiss. Berlin (Math. Phys. )} \textbf{1916}, 189

\bibitem{Schwarzschild:1916ae}
Schwarzschild~K 1916 \textit{
Sitzungsber. Preuss. Akad. Wiss. Berlin (Math. Phys. )} \textbf{1916}, 424

\bibitem{Tolman:1939jz}
Tolman~R~C 1939 \textit{
Phys. Rev.} \textbf{55}, 364

\bibitem{Oppenheimer:1939ne}
Oppenheimer~J~R and Volkoff~G~M 1939 \textit{
Phys. Rev.} \textbf{55}, 374

\bibitem{Kerr:1963ud}
Kerr~R~P 1963 \textit{
Phys. Rev. Lett.} \textbf{11}, 237

\bibitem{Boyer:1967}
Boyer~R~H and Lindquist~R~W 1967 \textit{
J. Math. Phys.} \textbf{8} (2), 265

\bibitem{Chandrasekhar:1985kt}
Chandrasekhar~S (\textit{The Mathematical Theory of Black Holes}, Oxford Press, 1985)
ISBN: 9780198503705.


\bibitem{Newman:1965tw}
Newman~E~T and Janis~A~I 1965 \textit{
J. Math. Phys.} \textbf{6}, 915

\bibitem{Newman:1965my}
Newman~E~T, Couch~E, Chinnapared~K, Exton~A, Prakash~A and Torrence~R 1965 \textit{
J. Math. Phys.} \textbf{6}, 918

\bibitem{Poisson:2004}
Poisson~E (\textit{A Relativist's Toolkit: The Mathematics of Black-Hole Mechanics}, Cambridge: Cambridge University Press, 2004. Section 4.3.5 and Section 5.1.8)

\bibitem{Griffiths:2009}
Griffiths~J~B and Podolsky~J (\textit{Exact Space-Times in Einstein's General Relativity}, Cambridge: Cambridge University Press, 2009. Section 9.5)


\bibitem{Friedmann:1924bb}
Friedmann~A 1924 \textit{
Z. Phys.} \textbf{21}, 326

\bibitem{Lemaitre:1927zz}
Lemaitre~G 1927 \textit{
Annales Soc. Sci. Bruxelles A} \textbf{47}, 49

\bibitem{Robertson:1933zz}
Robertson~H~P 1933 \textit{
Rev. Mod. Phys.} \textbf{5}, 62

\bibitem{Walker:1937qxv}
Walker~A~G 1937 \textit{
Proc. Lond. Math. Soc. s} \textbf{2-42}, no.1, 90

\bibitem{Penrose:1964wq}
Penrose~P 1965 \textit{
Phys. Rev. Lett.} \textbf{14} 57 

\bibitem{Hartle:1967he}
Hartle~J~B 1967 \textit{
Astrophys. J.} \textbf{150}, 1005

\bibitem{Hartle:1968si}
Hartle~J~B and Thorne~K~S 1968 \textit{
Astrophys. J.} \textbf{153}, 807

\bibitem{Regge:1957td}
Regge~T and Wheeler~J~A 1957 \textit{
Phys. Rev.} \textbf{108}, 1063

\bibitem{Thorne:1997kt}
Thorne~K~S 1998 \textit{
Phys. Rev. D} \textbf{58}, 124031

\bibitem{Arnowitt:1962hi}
Arnowitt~R~L, Deser~S and Misner~C~W 2008 \textit{
Gen. Rel. Grav.} \textbf{40}, 1997

\bibitem{Belinski:1971}
Belinski~V~A, Khalatnikov~I~M, and Lifshitz~E~M 1971 \textit{
Sov.Phys.Usp.} \textbf{13} 745

\bibitem{Belinskii:1972sg}
Belinski~V~A, Lifshitz~E~M and Khalatnikov~M~I 1972 \textit{
Zh. Eksp. Teor. Fiz.} \textbf{62}, no.5, 1606
[erratum: 1975 \textit{Zh. Eksp. Teor. Fiz.} \textbf{68}, no.5, 1968; erratum: 1975 \textit{Sov. Phys. JETP} \textbf{41}, no.5, 985]

\bibitem{Belinskii:1973sud}
Belinski~V~A and Khalatnikov~I~M 1973 \textit{
Sov. Phys. JETP} \textbf{36}, 591

\bibitem{BK}
Belinski~V~A and Khalatnikov~I~M (\textit{On the influence of the spinor and electromagnetic fields on the cosmological singularity}, Torino: Rend. Sem. Mat. Univ. e Politec. 35:159-180)

\bibitem{Belinsky:1982pk}
Belinski~V~A, Khalatnikov~I~M and Lifshitz~E~M 1982 \textit{
Adv. Phys.} \textbf{31}, 639

\bibitem{Belinski:2017fas}
Belinski~V and Henneaux~M (\textit{The Cosmological Singularity}, Cambridge: Cambridge University Press, 2017)


\bibitem{Blanchet:2013haa}
Blanchet~L 2014 \textit{
Living Rev. Rel.} \textbf{17}, 2

\bibitem{Damour:2016gwp}
Damour~T 2016 \textit{
Phys. Rev. D} \textbf{94}, no.10, 104015


\bibitem{Buonanno:1998gg}
Buonanno~A and Damour~T 1999 \textit{
Phys. Rev. D} \textbf{59}, 084006


\bibitem{Damour:2009zoi}
Damour~T and Nagar~A 2011 \textit{
Fundam. Theor. Phys.} \textbf{162}, 211

\bibitem{Damour:2012mv}
Damour~T 2014 \textit{
Fundam. Theor. Phys.} \textbf{177}, 111


\bibitem{Chow:2004}
Chow~B and Knopf~D (\textit{The Ricci Flow: An Introduction}, Providence, R.I.: American Mathematical Society, 2004)
ISBN 0-8218-3515-7.

\bibitem{MacKay:2024qxj}
MacKay~N~M 2025 \textit{
Class. Quant. Grav.} \textbf{42}, no.24, 245003

\bibitem{MacKay:2025uyg}
MacKay~N~M 2026 \textit{Class. Quant. Grav.}, in press~ doi:10.1088/1361-6382/ae72e4

\bibitem{GWOSC}
Gravitational Wave Open Science Center ({https://gwosc.org/eventapi/html/GWTC/})

\bibitem{LIGOScientific:2018mvr}
Abbott~B~P, LIGO Scientific and Virgo,~ \textit{et al.} 2019 \textit{
Phys. Rev. X} \textbf{9}, no.3, 031040

\bibitem{LIGOScientific:2021usb}
Abbott~R, LIGO Scientific and VIRGO,~\textit{et al.} 2024 \textit{
Phys. Rev. D} \textbf{109}, no.2, 022001

\bibitem{KAGRA:2021vkt}
Abbott~R, KAGRA, VIRGO and LIGO Scientific, \textit{et al.} 2023 \textit{
Phys. Rev. X} \textbf{13}, no.4, 041039



\bibitem{LIGOScientific:2025slb}
Abac~A~G, LIGO Scientific, VIRGO and KAGRA,~\textit{et al.} 
[arXiv:2508.18082 [gr-qc]].


\bibitem{Eddington:1924pmh}
Eddington~A~S 1924 \textit{
Nature} \textbf{113}, no.2832, 192 

\bibitem{Finkelstein:1958zz}
Finkelstein~D 1958 \textit{
Phys. Rev.} \textbf{110}, 965


\bibitem{Hawking:1974rv}
Hawking~S~W 1974 \textit{
Nature} \textbf{248}, 30

\bibitem{Hawking:1975vcx}
Hawking~S~W 1975 \textit{
Commun. Math. Phys.} \textbf{43}, 199 
[erratum: 1976 \textit{Commun. Math. Phys.} \textbf{46}, 206]

\bibitem{dInverno:1992gxs}
d'Inverno~R 1992


\bibitem{Miller:2025yyx}
Miller~A~L
[arXiv:2503.02607 [astro-ph.HE]].

\bibitem{Amaro-Seoane:2012aqc}
Amaro-Seoane~P, Aoudia~S, Babak~S, Binetruy~P, Berti~E, Bohe~A, Caprini~C, Colpi~M, Cornish~N~J and Danzmann~K, \textit{et al.} 2013
\textit{GW Notes} \textbf{6}, 4

\bibitem{Abdolrahimi:2016emo}
Abdolrahimi~S, Page~D~N and Tzounis~C 2019 \textit{
Phys. Rev. D} \textbf{100}, no.12, 124038

\bibitem{Hawking:1971ei}
Hawking~S 1971 \textit{
Mon. Not. Roy. Astron. Soc.} \textbf{152}, 75

\bibitem{Carr:2005qn}
Carr~B~J and Giddings~S~B 2005 \textit{
Sci. Am.} \textbf{292N5}, 30

\bibitem{Page:1976df}
Page~D~N 1976 \textit{
Phys. Rev. D} \textbf{13} 198

\bibitem{Page:1976ki}
Page~D~N 1976 \textit{
Phys. Rev. D} \textbf{14}, 3260

\bibitem{Jensen:1990xbb}
Jensen~B~P, McLaughlin~J and Ottewill~A~C 1991 \textit{
Phys. Rev. D} \textbf{43}, 4142

\bibitem{MacKay:2025yja}
MacKay~N~M 2025 \textit{
Phys. Lett. B} \textbf{870}, 139888

\bibitem{Oppenheimer:1939ue}
Oppenheimer~J~R and Snyder~H 1939 \textit{
Phys. Rev.} \textbf{56}, 455




\bibitem{Birrell:1982ix}
Birrell~N~D and Davies~P~C~W (\textit{Quantum Fields in Curved Space} Cambridge: Cambridge University Press, 1982)
ISBN: 978-0-511-62263-2, 978-0-521-27858-4

\bibitem{Choquet-Bruhat:2014}
Choquet-Bruhat~Y ('The Schwarzschild spacetime', \textit{Introduction to General Relativity, Black Holes, and Cosmology}, Oxford, 2014; online edn, Oxford Academic, 2024)

\bibitem{Bunch:1979uk}
Bunch~T~S and Parker~L 1979 \textit{
Phys. Rev. D} \textbf{20}, 2499

\bibitem{Bekenstein:1981xe}
Bekenstein~J~D and Parker~L 1981 \textit{
Phys. Rev. D} \textbf{23}, 2850

\bibitem{Alsing:2000ji}
Alsing~P~M, Evans~J~C and Nandi~K~K 2001 \textit{
Gen. Rel. Grav.} \textbf{33}, 1459

\bibitem{Li:2024ltx}
Li~B
[arXiv:2411.15164 [physics.gen-ph]].

\bibitem{MacKay:2025dbu}
MacKay~N~M
[arXiv:2501.15672 [physics.gen-ph]].


\bibitem{Gamow:1946hyv}
Gamow~G 1946 \textit{
Nature} \textbf{158}, no.4016, 549

\bibitem{Godel:1949ga}
G\"odel~K 1949 \textit{
Rev. Mod. Phys.} \textbf{21}, 447

\bibitem{Su:2009fu}
Su~S~C and Chu~M~C 2009 \textit{
Astrophys. J.} \textbf{703}, 354


\bibitem{Brevik:2014cxa}
Brevik~I and Gr{\o}n~{\O}
[arXiv:1409.8561 [gr-qc]].


\bibitem{Choudhuri:2015}
Choudhuri~A 2015 
\textit{Phys. Scr.} \textbf{90} 055004

\bibitem{Srivastava:2018gye}
Srivastava~M and Singh~C~P 2018 \textit{
Astrophys. Space Sci.} \textbf{363}, no.6, 117

\bibitem{Hu:2020xus}
Hu~J and Hu~H 2020 \textit{
Eur. Phys. J. Plus} \textbf{135}, no.9, 718 


\bibitem{Arora:2021tuh}
Arora~S, Pacif~S~K~J, Parida~A and Sahoo~P~K 2022 \textit{
JHEAp} \textbf{33}, 1

\bibitem{Singh:2023evc}
Singh~S~S, Kumrah~L, Alam~M~K, Singh~L~K and Devi~L~A 2024 \textit{
Can. J. Phys.} \textbf{102}, no.1, 61 


\bibitem{Hayden:2007cs}
Hayden~P and Preskill~J 2007 \textit{
JHEP} \textbf{09} 120

\bibitem{Manikandan:2018urq}
Manikandan~S~K and Jordan~A~N 2018 \textit{
Phys. Rev. D} \textbf{98}, no.12 124043

\bibitem{Calmet:2017hja}
Calmet~X, Kuntz~I and Moss~I~G 2018 \textit{
Found. Phys.} \textbf{48}, no.1, 110

\bibitem{Steinwachs:2019hdr}
Steinwachs~C~F 2020 \textit{
Fundam. Theor. Phys.} \textbf{199}, 253

\bibitem{Sahni:1999qe}
Sahni~V and Wang~L~M 2000 \textit{
Phys. Rev. D} \textbf{62}, 103517

\bibitem{Matos:2000ng}
Matos~T and Urena-Lopez~L~A 2000 \textit{
Class. Quant. Grav.} \textbf{17}, L75

\bibitem{Matos:2000ss}
Matos~T and Urena-Lopez~L~A 2001 \textit{
Phys. Rev. D} \textbf{63}, 063506

\bibitem{Alcubierre:2001ea}
Alcubierre~M, Guzman~F~S, Matos~T, Nunez~D, Urena-Lopez~L~A and Wiederhold~P 2002 \textit{
Class. Quant. Grav.} \textbf{19}, 5017

\bibitem{Bernal:2006it}
Bernal~A and Guzman~F~S 2006 \textit{
Phys. Rev. D} \textbf{74}, 063504

\bibitem{Boehm:2020wbt}
Boehm~C, Chu~X, Kuo~J~L and Pradler~J 2021 \textit{
Phys. Rev. D} \textbf{103}, no.7, 075005


\end{thebibliography}
\end{document}